\documentclass[12pt]{iopart}
\usepackage{iopams}  
\usepackage{graphicx}

  \expandafter\let\csname equation*\endcsname\relax
  \expandafter\let\csname endequation*\endcsname\relax

\usepackage{amsmath,amsfonts,amssymb}
\usepackage[margin=1in]{geometry}
\usepackage{array}
\usepackage{caption}
\usepackage{tikz}
\usepackage{listings}
\usetikzlibrary{snakes}
\usetikzlibrary{plotmarks}
\usepackage{bm}   
\usepackage{cite} 

\usepackage{float}
\usepackage{latexsym,mathrsfs}
\usepackage{epstopdf}
\usepackage{cite} 

\def \uqsl2 {$U_q$(sl$_2$) }
\def \cA {{\cal A}}

\newcommand {\be}{\begin{equation}}
\newcommand {\ee} {\end{equation}}
\newcommand {\bea}{\begin{eqnarray}}
\newcommand {\eea} {\end{eqnarray}}

\newcommand{\RI}{
\begin{tikzpicture}[scale=0.5,rotate=45]
\draw [black, line width=1pt, dashed] (-1,-1) -- (1,1); 
\draw [black, line width=1pt, dashed] (1,-1) -- (-1,1); 
\end{tikzpicture}}
\newcommand{\RIeightv}{
\begin{tikzpicture}[scale=0.701]
\draw [black, line width=0.1] (0,-1) -- (0,1); 
\draw [black, line width=0.1] (-1,0) -- (1,0); 
\draw[black, line width=1pt,<-,>=latex] (0,0.35) -- (0,0.45);
\draw[black, line width=1pt,<-,>=latex] (0,-0.65) -- (0,-0.55);
\draw[black, line width=1pt,<-,>=latex] (0.35,0) -- (0.45,0);
\draw[black, line width=1pt,<-,>=latex] (-0.65,0) -- (-0.55,0);
\end{tikzpicture}}
\newcommand{\RII}{
\begin{tikzpicture}[scale=0.5,rotate=45]
\draw [black, line width=1pt, dashed] (-1,-1) -- (1,1); 
\draw [black, line width=1pt, dashed] (1,-1) -- (-1,1); 
\draw[red, line width=2pt, rounded corners=7pt] (-1,-1) -- (0,0) -- (-1,1);
\end{tikzpicture}}
\newcommand{\RIIeightv}{
\begin{tikzpicture}[scale=0.701]
\draw [black, line width=0.1] (0,-1) -- (0,1); 
\draw [black, line width=0.1] (-1,0) -- (1,0); 
\draw[black, line width=1pt,<-,>=latex] (0,0.35) -- (0,0.45);
\draw[black, line width=1pt,->,>=latex] (0,-0.45) -- (0,-0.35);
\draw[black, line width=1pt,<-,>=latex] (0.35,0) -- (0.45,0);
\draw[black, line width=1pt,->,>=latex] (-0.45,0) -- (-0.35,0);
\end{tikzpicture}}
\newcommand{\RIII}{
\begin{tikzpicture}[scale=0.5,rotate=45]
\draw [black, line width=1pt, dashed] (-1,-1) -- (1,1); 
\draw [black, line width=1pt, dashed] (1,-1) -- (-1,1);  
\draw[red, line width=2pt, rounded corners=7pt] (1,-1) -- (0,0) -- (1,1);
\end{tikzpicture}}
\newcommand{\RIIIeightv}{
\begin{tikzpicture}[scale=0.701]
\draw [black, line width=0.1] (0,-1) -- (0,1); 
\draw [black, line width=0.1] (-1,0) -- (1,0); 
\draw[black, line width=1pt,->,>=latex] (0,0.55) -- (0,0.65);
\draw[black, line width=1pt,<-,>=latex] (0,-0.65) -- (0,-0.55);
\draw[black, line width=1pt,->,>=latex] (0.55,0) -- (0.65,0);
\draw[black, line width=1pt,<-,>=latex] (-0.65,0) -- (-0.55,0);
\end{tikzpicture}}
\newcommand{\RIV}{
\begin{tikzpicture}[scale=0.5,rotate=45]
\draw [black, line width=1pt, dashed] (-1,-1) -- (1,1); 
\draw [black, line width=1pt, dashed] (1,-1) -- (-1,1); 
\draw[red, line width=2pt, rounded corners=7pt] (-1,-1) -- (0,0) -- (1,-1);
\end{tikzpicture}}
\newcommand{\RIVeightv}{
\begin{tikzpicture}[scale=0.701]
\draw [black, line width=0.1] (0,-1) -- (0,1); 
\draw [black, line width=0.1] (-1,0) -- (1,0); 
\draw[black, line width=1pt,<-,>=latex] (0,0.35) -- (0,0.45);
\draw[black, line width=1pt,->,>=latex] (0,-0.45) -- (0,-0.35);
\draw[black, line width=1pt,->,>=latex] (0.55,0) -- (0.65,0);
\draw[black, line width=1pt,<-,>=latex] (-0.65,0) -- (-0.55,0);
\end{tikzpicture}}
\newcommand{\RV}{
\begin{tikzpicture}[scale=0.5,rotate=45]
\draw [black, line width=1pt, dashed] (-1,-1) -- (1,1); 
\draw [black, line width=1pt, dashed] (1,-1) -- (-1,1);  
\draw[red, line width=2pt, rounded corners=7pt] (-1,1) -- (0,0) -- (1,1);
\end{tikzpicture}}
\newcommand{\RVeightv}{
\begin{tikzpicture}[scale=0.701]
\draw [black, line width=0.1] (0,-1) -- (0,1); 
\draw [black, line width=0.1] (-1,0) -- (1,0); 
\draw[black, line width=1pt,->,>=latex] (0,0.55) -- (0,0.65);
\draw[black, line width=1pt,<-,>=latex] (0,-0.65) -- (0,-0.55);
\draw[black, line width=1pt,<-,>=latex] (0.35,0) -- (0.45,0);
\draw[black, line width=1pt,->,>=latex] (-0.45,0) -- (-0.35,0);
\end{tikzpicture}}

\newcommand{\RVIII}{
\begin{tikzpicture}[scale=0.5,rotate=45]
\draw [black, line width=1pt, dashed] (-1,-1) -- (1,1); 
\draw [black, line width=1pt, dashed] (1,-1) -- (-1,1); 
\draw[red, line width=2pt, rounded corners=7pt] (-1,-1) -- (0,0) -- (-1,1);
\draw[red, line width=2pt, rounded corners=7pt] (1,-1) -- (0,0) -- (1,1);
\end{tikzpicture}}
\newcommand{\RVIIIeightv}{
\begin{tikzpicture}[scale=0.701]
\draw [black, line width=0.1] (0,-1) -- (0,1); 
\draw [black, line width=0.1] (-1,0) -- (1,0); 
\draw[black, line width=1pt,->,>=latex] (0,0.55) -- (0,0.65);
\draw[black, line width=1pt,->,>=latex] (0,-0.45) -- (0,-0.35);
\draw[black, line width=1pt,->,>=latex] (0.55,0) -- (0.65,0);
\draw[black, line width=1pt,->,>=latex] (-0.45,0) -- (-0.35,0);
\end{tikzpicture}}
\newcommand{\RIX}{
\begin{tikzpicture}[scale=0.5,rotate=45]
\draw [black, line width=1pt, dashed] (-1,-1) -- (1,1); 
\draw [black, line width=1pt, dashed] (1,-1) -- (-1,1);   
\draw[red, line width=2pt, rounded corners=7pt] (-1,-1) -- (0,0) -- (1,-1);
\draw[red, line width=2pt, rounded corners=7pt] (-1,1) -- (0,0) -- (1,1);
\end{tikzpicture}}
\newcommand{\RIXeightv}{
\begin{tikzpicture}[scale=0.701]
\draw [black, line width=0.1] (0,-1) -- (0,1); 
\draw [black, line width=0.1] (-1,0) -- (1,0); 
\draw[black, line width=1pt,->,>=latex] (0,0.55) -- (0,0.65);
\draw[black, line width=1pt,->,>=latex] (0,-0.45) -- (0,-0.35);
\draw[black, line width=1pt,->,>=latex] (0.55,0) -- (0.65,0);
\draw[black, line width=1pt,->,>=latex] (-0.45,0) -- (-0.35,0);
\end{tikzpicture}}

\begin{document}

\title[Dilute oriented loop models]{Dilute oriented loop models}

\author{Eric Vernier$^{1,2,3}$, Jesper Lykke Jacobsen$^{1,3}$, and Hubert Saleur$^{2,4}$}

\address{${}^1$LPTENS, \'Ecole Normale Sup\'erieure -- 
PSL Research University, 24 rue Lhomond, F-75231 Paris Cedex 05, France}
\address{${}^2$Institut de Physique Th\'eorique, CEA Saclay, 91191 
Gif Sur Yvette, France}
\address{${}^3$Sorbonne Universit\'es, UPMC Universit\'e Paris 6, 
CNRS UMR 8549, F-75005 Paris, France} 
\address{${}^4$USC Physics Department, Los Angeles CA 90089, USA}

\eads{\mailto{ejvernier@gmail.com}, \mailto{jesper.jacobsen@ens.fr}, 
      \mailto{hubert.saleur@cea.fr}}

\begin{abstract}

We study a model of dilute oriented loops on the square lattice, where each loop is compatible with a fixed, alternating orientation of the lattice edges.
This implies that loop strands are not allowed to go straight at vertices, and results in an enhancement of the usual ${\rm O}(n)$ symmetry to ${\rm U}(n)$.
The corresponding transfer matrix acts on a number of representations (standard modules) that grows exponentially with the system size.
We derive their dimension and those of the centraliser by both combinatorial and algebraic techniques. A mapping onto a field theory
permits us to identify the conformal field theory governing the critical range, $n \le 1$. We establish the phase diagram and the critical
exponents of low-energy excitations. For generic $n$, there is a critical line in the universality class of the dilute ${\rm O}(2n)$ model, terminating
in an ${\rm SU}(n+1)$ point. The case $n=1$ maps onto the critical line of the six-vertex model, along which exponents vary continuously.

\end{abstract}

\section{Introduction}

\begin{figure}[H]
\centering
\begin{tikzpicture}[scale=0.75]
\foreach \x in {0,2,...,4}  
{ 
\foreach \y in {0,2,...,4}
{ 
\draw[black,line width=1.5pt,->,>=latex] (\x,\y) -- (\x+1,\y+1);
\draw[black,line width=1.5pt,->,>=latex]  (\x+1,\y+1)--(\x,\y+2);
\draw[black,line width=1.5pt,->,>=latex] (\x+2,\y+2) -- (\x+1,\y+1);
\draw[black,line width=1.5pt,->,>=latex]  (\x+1,\y+1)--(\x+2,\y) ;
}}
\end{tikzpicture}  
\hspace{2cm}
\begin{tikzpicture}[scale=0.75]
\draw[red,line width=3.5pt, rounded corners=10pt] (0,0) -- (1,1) -- (2,0)-- (3,1)-- (2,2)-- (3,3)-- (2,4)-- (3,5)-- (4,4)-- (5,5)-- (4.1,5.9);
\draw[red,line width=3.5pt, rounded corners=10pt] (0.5,1.5) -- (1,1)-- (2,2)-- (1,3)-- (2,4)-- (1,5)-- (0,4)-- (1,3)-- (0,2)-- (0.5,1.5);
\draw[red,line width=3.5pt, rounded corners=10pt] (0.,6) -- (1,5) -- (2,6) -- (3,5) -- (3.9,5.9);
\draw[red,line width=3.5pt, rounded corners=10pt] (6.,6)-- (5,5)-- (6,4)-- (5,3)-- (4,4)-- (3,3)-- (4,2)-- (5,3)-- (6,2)-- (5,1)-- (6,0);
\draw[red,line width=3.5pt, rounded corners=10pt] (4.5,0.5)-- (5,1)-- (4,2)-- (3,1)-- (4,0)-- (4.5,0.5);
   \foreach \x in {0,2,...,4}  
{ 
\foreach \y in {0,2,...,4}
{ 
\draw[black,line width=1.0pt,->,>=latex] (\x,\y) -- (\x+1,\y+1);
\draw[black,line width=1.0pt,->,>=latex]  (\x+1,\y+1)--(\x,\y+2);
\draw[black,line width=1.0pt,->,>=latex] (\x+2,\y+2) -- (\x+1,\y+1);
\draw[black,line width=1.0pt,->,>=latex]  (\x+1,\y+1)--(\x+2,\y) ;
}}
\end{tikzpicture}  
  \caption{Left panel: oriented square lattice used for the definition of the various loop models in this paper. Right panel: configuration of completely packed, self-avoiding and mutually avoiding loops on this oriented lattice.}
 \label{figdense}
\end{figure}
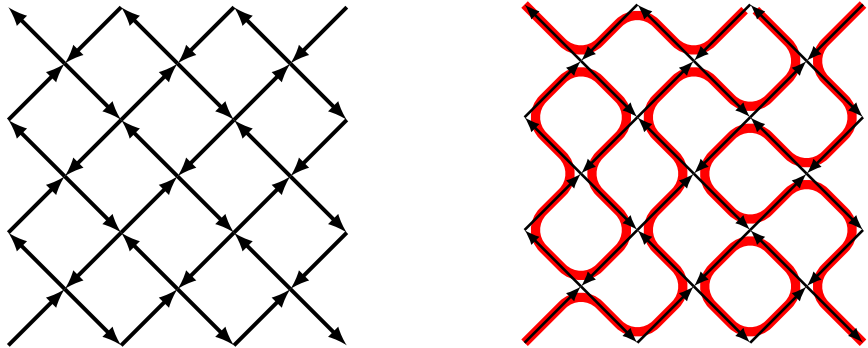

Loop models and their critical universality classes play a major role in several areas of theoretical physics, such as the study of geometrical statistical models,  quantum integrable models, conformal field theory \cite{JacobsenReview}, or quantum information theory \cite{JS-valence-bond}. The simplest---and most studied---of all the loop models is conveniently described by starting from an oriented square lattice, represented in figure \ref{figdense}.
Every edge of the lattice is then supposed to be occupied by a loop segment (a monomer), while  vertices can be split in two possible ways compatible with the link orientations. This gives rise to configurations of completely packed, self-avoiding and mutually avoiding loops (see figure \ref{figdense}). Finally, a loop fugacity $n$ is introduced, giving to each configuration a weight $n^{N_{\rm loops}}$, where $N_{\rm loops}$ is the number of loops. 

It is well-known that this model has an underlying ${\rm U}(n)$ symmetry when $n$ is an integer \cite{Affleck90,ReadSaleur07}.
This can be seen in two ways. The first stems from calculating the partition function of this model by a transfer matrix approach, with the $x$-axis representing the `space' and the $y$-axis the `imaginary time' direction. The transfer matrix is then a product of elementary operators at every vertex which produce the two possible splits that respect the lattice orientation.
Graphically, each vertex can be represented as follows:
\begin{center}
\begin{tikzpicture}
\begin{scope}[shift={(-2,0)}]
\draw[black,line width=1.5pt,<-,>=latex] (0,0) -- (0.5,0.5);
\draw[black,line width=1.5pt,<-,>=latex]  (0.5,0.5)--(0,1);
\draw[black,line width=1.5pt,<-,>=latex] (1,1) -- (0.5,0.5);
\draw[black,line width=1.5pt,<-,>=latex]  (0.5,0.5)--(1,0) ;
\end{scope}
\node at (-0.5,0.) {$,$};
\draw[black,line width=1.5pt,->,>=latex] (0,0) -- (0.5,0.5);
\draw[black,line width=1.5pt,->,>=latex]  (0.5,0.5)--(0,1);
\draw[black,line width=1.5pt,->,>=latex] (1,1) -- (0.5,0.5);
\draw[black,line width=1.5pt,->,>=latex]  (0.5,0.5)--(1,0) ;
\node at (2,0.5) {$=$};
\begin{scope}[shift={(3,0)}]
\draw(0,0)--(1,1);
\draw(0,1)--(1,0);
\draw[red,line width=2pt,rounded corners=8pt] (0,0) -- (0.5,0.5) -- (0,1);
\draw[red,line width=2pt,rounded corners=8pt] (1,0) -- (0.5,0.5) -- (1,1);
\end{scope}
\node at (5,0.5) {$+$};
\node at (5.7,0.5) {$x$};
\begin{scope}[shift={(6,0)}]
\draw(0,0)--(1,1);
\draw(0,1)--(1,0);
\draw[red,line width=2pt,rounded corners=8pt] (1,1) -- (0.5,0.5) -- (0,1);
\draw[red,line width=2pt,rounded corners=8pt] (0,0) -- (0.5,0.5) -- (1,0);
\end{scope}
\end{tikzpicture}
\end{center}
or, symbolically,
\begin{equation}
t = I  + x e \,,
\label{eq:vertexTL}
\end{equation}
where the $e_i$ are generators of the Temperley Lieb (TL) algebra, acting on strands number $i$ and $i+1$.
They  satisfy a well-known set of multiplication rules  that can be read from  their geometrical definition \cite{TemperleyLieb71,Baxter_book}. In particular, 
 
\begin{equation}
\left(e_{i}\right)^2 =
 \begin{tikzpicture}[baseline={([yshift=0.9ex]current bounding box.center)}]
   \foreach \x in {0,0.25,0.5,1.25,2,2.75,3}{\draw[red, line width=0.6mm, rounded corners=7pt] (\x,-0.25) -- (\x,0.25);}
   \begin{scope}[shift={(0.25,0)},rotate=0] 
   \draw[red, line width=0.6mm] (1.25,-0.25) arc(180:0:0.125 and 0.2);
   \draw[red, line width=0.6mm] (1.25,0.25) arc(180:360:0.125 and 0.2);
   \end{scope}
   \node at (0.9,0) {$\ldots$};
    \node at (2.4,0) {$\ldots$};
    \node[below] at (0,-0.25) {$1$};
    \node[below] at (0.25,-0.25) {$2$};
      \node[below] at (0.5,-0.25) {$3$};
\node[below] at (1.5,-0.25) {$i$};
    \node[below] at (3,-0.25) {$N$};

\begin{scope}[shift={(0,0.5)},rotate=0] 
   \foreach \x in {0,0.25,0.5,1.25,2,2.75,3}{\draw[red, line width=0.6mm, rounded corners=7pt] (\x,-0.25) -- (\x,0.25);}
   \begin{scope}[shift={(0.25,0)},rotate=0] 
   \draw[red, line width=0.6mm] (1.25,-0.25) arc(180:0:0.125 and 0.2);
   \draw[red, line width=0.6mm] (1.25,0.25) arc(180:360:0.125 and 0.2);
\   \end{scope}
   \end{scope}
\end{tikzpicture}
= n \,
 \begin{tikzpicture}[baseline={([yshift=0.9ex]current bounding box.center)}]
   \foreach \x in {0,0.25,0.5,1.25,2,2.75,3}{\draw[red, line width=0.6mm, rounded corners=7pt] (\x,-0.25) -- (\x,0.25);}
   \begin{scope}[shift={(0.25,0)},rotate=0] 
   \draw[red, line width=0.6mm] (1.25,-0.25) arc(180:0:0.125 and 0.2);
   \end{scope}
   \node at (0.9,0) {$\ldots$};
    \node at (2.4,0) {$\ldots$};
    \node[below] at (0,-0.25) {$1$};
    \node[below] at (0.25,-0.25) {$2$};
      \node[below] at (0.5,-0.25) {$3$};
\node[below] at (1.5,-0.25) {$i$};
    \node[below] at (3,-0.25) {$N$};

\begin{scope}[shift={(0,0.5)},rotate=0] 
   \foreach \x in {0,0.25,0.5,1.25,2,2.75,3}{\draw[red, line width=0.6mm, rounded corners=7pt] (\x,-0.25) -- (\x,0.25);}
   \begin{scope}[shift={(0.25,0)},rotate=0] 
   \draw[red, line width=0.6mm] (1.25,0.25) arc(180:360:0.125 and 0.2);
   \end{scope}
   \end{scope}
\end{tikzpicture}
 = n \, e_{i} \,,
 \label{eq:TL:1}
\end{equation}
so as to give every loop a weight $n$. The parity of the number of lattice sites $N$ will play some role in the following.
In the case of periodic boundary conditions horizontally, the consistency of the orientations of edges (see figure~\ref{figdense})
requires $N$ to be even. With free transverse boundary conditions $N$ can have any parity, but in most of the paper (unless
explicitly stating the contrary) we shall nevertheless set $N = 2L$ also in that case.

The TL algebra admits a simple Hilbert space realisation where links oriented to the left (resp.\ the right) on a horizontal line carry a fundamental $\mathfrak{n}$ (resp.\ conjugate fundamental $\bar{\mathfrak{n}}$) representation of ${\rm U}(n)$.%
\footnote{The fundamental representation $\mathfrak{n}$ should not be confused with the loop weight $n \in \mathbb{C}$. It is mathematically well-defined only
for $n \in \mathbb{N}$, in which case the dimension of $\mathfrak{n}$ is $n$. However, most results throughout this paper make sense by analytic continuation
for arbitrary $n \in \mathbb{R}$.}
The generators $e_i$ can then be written as $e_i=n P_{}$, where $P$ is the projector onto the identity in the product $\mathfrak{n} \otimes \bar{\mathfrak{n}}$ (resp.\  $\bar{\mathfrak{n}} \otimes \mathfrak{n}$). The transfer matrix (or, in the anisotropic limit, the Hamiltonian) then acquires a ${\rm U}(n)$ symmetry. 

General arguments mapping the spin chain Hamiltonian to a sigma model (see, e.g., \cite{Affleck85,ReadSachdev89}) allow one to identify the long-distance physics of the loop model with that of the ${\rm CP}^{n-1}$ sigma model with a topological angle  $\theta=\pi$. The loop model sits at a point of first-order phase transition for $n>2$, and is critical for $n\in [-2,2]$. It is widely believed that the sigma model enjoys the same properties, and in fact, the loop model was historically studied in part because of the correspondence with ${\rm CP}^{n-1}$ \cite{Affleck85}. 

Another, maybe more physical, way to see the emergence of the ${\rm U}(n)$ symmetry (see \cite{Nahum11,Nahum13} for closely related work in both three and two dimensions) is to consider directly the Euclidean version, and observe that  the partition function can be calculated  by introducing $n$-dimensional complex vectors $\vec{z}$ that live on the edges,%
\footnote{One may naively wonder why a {\sl complex} vector $\vec{z}$ is needed for the loop model. The point is that we need the interaction to decompose into two diagrams only, and this is what happens in $\mathfrak{n} \otimes \bar{\mathfrak{n}}$ in ${\rm U}(n)$. In contrast, in ${\rm O}(n)$, we have three diagrams. This means that an interaction where we would replace the $\vec{z} \cdot \vec{z'}^\dagger$ by $\vec{n} \cdot \vec{n'}$  could not be interpreted unambiguously in terms of loops.}
with interactions that match the geometrical definition. This is done by associating, for instance, to the vertex

\begin{center}
\begin{tikzpicture}[scale=1.5]
\begin{scope}[shift={(-2,0)}]
\draw[black,line width=1.5pt,->,>=latex] (0,0) -- (0.5,0.5);
\draw[black,line width=1.5pt,->,>=latex]  (0.5,0.5)--(0,1);
\draw[black,line width=1.5pt,->,>=latex] (1,1) -- (0.5,0.5);
\draw[black,line width=1.5pt,->,>=latex]  (0.5,0.5)--(1,0) ;
\node[left] at (0.3,0.3) {$B$};
\node[left] at (0.3,0.7) {$A$};
\node[right] at (0.7,0.3) {$C$};
\node[right] at (0.7,0.7) {$D$};
\end{scope}
\end{tikzpicture}
\end{center}
a term
\begin{equation}
e^{-S}=\ldots \left[p(\vec{z}_A^\dagger \cdot \vec{z}_B)(\vec{z}_C^\dagger \cdot \vec{z}_D)+(1-p)
(\vec{z}_A^\dagger \cdot \vec{z}_D)(\vec{z}_C^\dagger \cdot \vec{z}_B)\right]\ldots\label{BWexp}
\end{equation}
where the vectors $\vec{z}$ and $\vec{z}^\dagger$, living respectively in the fundamental and conjugate fundamental representations, obey the normalisation $|\vec{z}|^2=|\vec{z}^\dagger|^2=1$, and the bracket represents the contribution to the Boltzmann weight from the corresponding vertex (note that there is no term without a vector $\vec{z}$, since the model is completely packed). The real parameter
$p$ can be interpreted as the probability of taking the first diagram in the expansion of (\ref{BWexp}). The partition function
\begin{equation}
Z\propto\prod_\text{edges $e$}\int {\rm d}\vec{z}_e \, e^{-S}
\end{equation}
can be expanded by picking either of the two terms in each vertex contribution,  giving rise immediately to the loop model with weight $n$ per loop, since the vectors $\vec{z}$ have $n$ components; the contribution to each vertex is actually proportional to that of (\ref{eq:vertexTL}), with $x=\frac{1-p}{p}$. Note that every $\vec{z}_e$ occurs {\sl twice},
since each edge is shared by a pair of vertices, and the contribution along a loop of length $k$ is of the form
\begin{equation*}
 {\rm Tr} \, (\vec{z}_{e_1}^\dagger \cdot \vec{z}_{e_2})(\vec{z}_{e_2}^\dagger \cdot \vec{z}_{e_3}) \cdots (\vec{z}_{e_k}^\dagger \cdot \vec{z}_{e_1}) \,.
\end{equation*}
This implies that the phase of $\vec{z}_e$ is not a physical degree of freedom, showing that the target is ${\rm U}(n)/{\rm U}(n-1)\times {\rm U}(1)$, that is, the projective space  ${\rm CP}^{n-1}$. We shall see below how the  interaction term of the sigma model can be obtained by taking the continuum limit of terms such as (\ref{BWexp}).  To see the origin of the topological term in this picture \cite{Affleck91}, one may argue as in \cite{Nahum-thesis} (see footnote 3 of chapter 3).

Various modifications of this model can be imagined, depending in particular on what happens to the symmetry. Of special interest is the loop model where {\sl crossings} are now allowed at the vertices \cite{MartinsNienhuisRietman98} and given some Boltzmann weights $w$, while the loop fugacity remains equal to $n$. Adding such an interaction breaks the ${\rm U}(n)$ symmetry down to ${\rm O}(n)$,
since the crossing `split' does not respect the lattice orientation.
The transfer matrix can then be written in terms of the projectors on the two independent representations that occur in the tensor product of the vector ${\rm O}(n)$ representation with itself. The universality class is modified, and has been shown  \cite{JacobsenReadSaleur03} to correspond to the low-temperature Goldstone phase of the ${\rm O}(n)$  model, which is described by the weak-coupling fixed point of the ${\rm O}(n)/{\rm O}(n-1)$ sigma model. 

Another familiar modification consists in diluting the loop model to allow edges that are not covered by loops \cite{BloteNienhuis89}. In general, this dilution is made in such a way that the ${\rm U}(n)$ symmetry is broken, and the loop trajectories do not respect the orientation of the lattice. The remaining symmetry is again only ${\rm O}(n)$ \cite{ReadSaleur07}. While the interaction is not the most general allowed by this symmetry (since no crossings occur), it is known that  crossings does not change further the universality class, which is generically ${\rm O}(n)$ criticality.

Meanwhile, a  modification preserving ${\rm U}(n)$ symmetry can be obtained  by allowing  next-nearest neighbour lines to cross, via the permutation generator ${\cal P}_{i,i+2}$. This is most elegantly studied on the triangular lattice, and gives interesting results in particular for $n=0$ \cite{CanduJacobsenReadSaleur10}. For other recent developments about loop models, see \cite{Nahum13}.

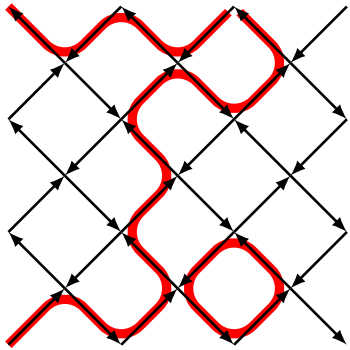
\begin{figure}
\centering
\begin{tikzpicture}[scale=0.75]
\draw[red,line width=3.5pt, rounded corners=10pt] (0,0) -- (1,1) -- (2,0)-- (3,1)-- (2,2)-- (3,3)-- (2,4)-- (3,5)-- (4,4)-- (5,5)-- (4.1,5.9);
\draw[red,line width=3.5pt, rounded corners=10pt] (0.,6) -- (1,5) -- (2,6) -- (3,5) -- (3.9,5.9);
\draw[red,line width=3.5pt, rounded corners=10pt] (4.5,0.5)-- (5,1)-- (4,2)-- (3,1)-- (4,0)-- (4.5,0.5);
   \foreach \x in {0,2,...,4}  
{ 
\foreach \y in {0,2,...,4}
{ 
\draw[black,line width=1.0pt,->,>=latex] (\x,\y) -- (\x+1,\y+1);
\draw[black,line width=1.0pt,->,>=latex]  (\x+1,\y+1)--(\x,\y+2);
\draw[black,line width=1.0pt,->,>=latex] (\x+2,\y+2) -- (\x+1,\y+1);
\draw[black,line width=1.0pt,->,>=latex]  (\x+1,\y+1)--(\x+2,\y) ;
}}
\end{tikzpicture}  
  \caption{Configuration of the oriented, dilute loop model.}
 \label{figdil}
\end{figure}

Another ${\rm U}(n)$-preserving modification that has not been studied much \cite{BloteNienhuis89} consists in {\sl diluting the loops while preserving the orientation of the underlying lattice} (see figure \ref{figdil}). 
This produces loops which never go straight, and that may or may not meet at vertices, with the  allowed configurations being:
\begin{equation}
\label{dildiags}
\begin{tikzpicture}
\begin{scope}[shift={(-1.6,0)}]
\draw[black,line width=1.5pt,<-,>=latex] (0,0) -- (0.5,0.5);
\draw[black,line width=1.5pt,<-,>=latex]  (0.5,0.5)--(0,1);
\draw[black,line width=1.5pt,<-,>=latex] (1,1) -- (0.5,0.5);
\draw[black,line width=1.5pt,<-,>=latex]  (0.5,0.5)--(1,0) ;
\end{scope}
\node at (-0.3,0.) {$,$};
\draw[black,line width=1.5pt,->,>=latex] (0,0) -- (0.5,0.5);
\draw[black,line width=1.5pt,->,>=latex]  (0.5,0.5)--(0,1);
\draw[black,line width=1.5pt,->,>=latex] (1,1) -- (0.5,0.5);
\draw[black,line width=1.5pt,->,>=latex]  (0.5,0.5)--(1,0) ;
\node at (2,0.5) {$=$};
\begin{scope}[shift={(3,0)}]
\draw(0,0)--(1,1);
\draw(0,1)--(1,0);
\end{scope}
\node at (5,0.5) {$+$};
\node at (5.7,0.5) {$K$};
\begin{scope}[shift={(6,0)}]
\draw(0,0)--(1,1);
\draw(0,1)--(1,0);
\draw[red,line width=2pt,rounded corners=8pt] (0,0) -- (0.5,0.5) -- (0,1);
\end{scope}
\node at (8,0.5) {$+$};
\node at (8.7,0.5) {$K$};
\begin{scope}[shift={(9,0)}]
\draw(0,0)--(1,1);
\draw(0,1)--(1,0);
\draw[red,line width=2pt,rounded corners=8pt] (1,0) -- (0.5,0.5) -- (1,1);
\end{scope}
\node at (11,0.5) {$+$};
\node at (11.7,0.5) {$K^2 \tau$};
\begin{scope}[shift={(12,0)}]
\draw(0,0)--(1,1);
\draw(0,1)--(1,0);
\draw[red,line width=2pt,rounded corners=8pt] (1,1) -- (0.5,0.5) -- (1,0);
\draw[red,line width=2pt,rounded corners=8pt] (0,0) -- (0.5,0.5) -- (0,1);
\end{scope}
\node at (5,-1) {$+$};
\node at (5.7,-1) {$K$};
\begin{scope}[shift={(6,-1.5)}]
\draw(0,0)--(1,1);
\draw(0,1)--(1,0);
\draw[red,line width=2pt,rounded corners=8pt] (1,1) -- (0.5,0.5) -- (0,1);
\end{scope}
\node at (8,-1) {$+$};
\node at (8.7,-1) {$K$};
\begin{scope}[shift={(9,-1.5)}]
\draw(0,0)--(1,1);
\draw(0,1)--(1,0);
\draw[red,line width=2pt,rounded corners=8pt] (0,0) -- (0.5,0.5) -- (1,0);
\end{scope}
\node at (11,-1) {$+$};
\node at (11.7,-1) {$K^2 \tau$};
\begin{scope}[shift={(12,-1.5)}]
\draw(0,0)--(1,1);
\draw(0,1)--(1,0);
\draw[red,line width=2pt,rounded corners=8pt] (1,1) -- (0.5,0.5) -- (0,1);
\draw[red,line width=2pt,rounded corners=8pt] (0,0) -- (0.5,0.5) -- (1,0);
\end{scope}
\end{tikzpicture}   
\end{equation}
The Boltzmann weights are here $K$ per monomer and $\tau$ per vertex encounter.
Our purpose in the following will be to study the critical properties of this model, which we shall call the `dilute oriented loop model'. 
In all the rest of this paper, we will put a tilde symbol ($\tilde{\ }$) on top of quantities refering to the usual (completely packed) TL model,
in order to distinguish them from quantities in the dilute oriented model.

The paper is organised as follows. In section~\ref{sec:TM} we discuss the transfer matrix of the dilute oriented model in terms
of its symmetries, the modules on which it acts, and its centraliser, both with free and periodic boundary conditions in the transverse
direction. We use to this end a mixture of combinatorial and algebraic techniques.
In section~\ref{sec:cft} we map the model onto a field theory. We discuss the low-energy limit of this theory and the corresponding critical
exponents. The phase diagram will be shown to contain a critical line where, remarkably, the symmetry is that of a dilute {\rm O}($2n$) model
(although the loop weight is just $n$). This line terminates in a point with ${\rm SU}(n+1)$ symmetry that is the dilute counterpart of the
completely packed model. The case $n=1$ is special, and we shall show that it contains the entire critical line of the six-vertex model,
along which the exponents vary continuously. While several types of open boundary conditions can be considered, we mostly focus our numerical study on the periodic case. However, the conclusions we draw are argued to be generic.  Finally, section~\ref{sec:disc} contains the discussion and a few concluding remarks.

\section{Transfer matrix and symmetries}
\label{sec:TM}

\subsection{The algebra}

Before discussing the possible universality classes of this model, it is interesting to study algebraic aspects of its transfer matrix description. This description naturally involves a `dilute' version of the Temperley-Lieb algebra, but not what is usually called \cite{GrimmPearce93,Grimm96} the `dilute Temperley-Lieb algebra', since in our model loop strands are never allowed to go straight at vertices. The implication for our dilute oriented loop model is that loop strands on odd sites can only be contracted with strands on even sites, just like in the completely packed case. 

The generators of the `dilute oriented Temperley-Lieb algebra' are the following 
 \begin{eqnarray}
e^+_{i} &=&
 \begin{tikzpicture}[baseline={([yshift=0.8ex]current bounding box.center)}]
   \foreach \x in {0,0.25,0.5,1.25,2,2.75,3}{\draw[red,dashed, line width=0.5mm, rounded corners=7pt] (\x,-0.25) -- (\x,0.25);}
   \begin{scope}[shift={(0.25,0)},rotate=0] 
   \draw[red, line width=0.6mm] (1.25,0.25) arc(180:360:0.125 and 0.2);
   \end{scope}
   \node at (0.9,0) {$\ldots$};
    \node at (2.4,0) {$\ldots$};
    \node[below] at (0,-0.25) {$1$};
    \node[below] at (0.25,-0.25) {$2$};
      \node[below] at (0.5,-0.25) {$3$};
\node[below] at (1.5,-0.25) {$i$};
    \node[below] at (3,-0.25) {$N$};
\end{tikzpicture}
 \label{eq:odTL}
 \\
e^-_{i} &=&
 \begin{tikzpicture}[baseline={([yshift=0.8ex]current bounding box.center)}]
   \foreach \x in {0,0.25,0.5,1.25,2,2.75,3}{\draw[red,dashed, line width=0.5mm, rounded corners=7pt] (\x,-0.25) -- (\x,0.25);}
   \begin{scope}[shift={(0.25,0)},rotate=0] 
   \draw[red, line width=0.6mm] (1.25,-0.25) arc(180:0:0.125 and 0.2);
   \end{scope}
   \node at (0.9,0) {$\ldots$};
    \node at (2.4,0) {$\ldots$};
    \node[below] at (0,-0.25) {$1$};
    \node[below] at (0.25,-0.25) {$2$};
      \node[below] at (0.5,-0.25) {$3$};
\node[below] at (1.5,-0.25) {$i$};
    \node[below] at (3,-0.25) {$N$};
\end{tikzpicture} \,,
\nonumber
\end{eqnarray}
where the dashed lines act as the identity, namely 
\begin{equation}
\begin{tikzpicture}
\draw[red,dashed, line width=0.5mm] (0,0) -- (0,0.5);
\node at (0.5,0.25) {$=$};
\draw (1,0) -- (1,0.5);
\node at (1.5,0.25) {$+$};
\draw[red,line width=0.5mm] (2,0) -- (2,0.5);
\end{tikzpicture}
\label{id-strand}
\end{equation}
Note that the dilute version of the usual TL generator [cf.~(\ref{eq:TL:1})] is then obtained as
\begin{equation}
e_{i} = e_i^+ e_i^- =
 \begin{tikzpicture}[baseline={([yshift=0.8ex]current bounding box.center)}]
   \foreach \x in {0,0.25,0.5,1.25,2,2.75,3}{\draw[red,dashed, line width=0.5mm, rounded corners=7pt] (\x,-0.25) -- (\x,0.25);}
   \begin{scope}[shift={(0.25,0)},rotate=0] 
   \draw[red, line width=0.6mm] (1.25,-0.25) arc(180:0:0.125 and 0.2);
   \draw[red, line width=0.6mm] (1.25,0.25) arc(180:360:0.125 and 0.2);
   \end{scope}
   \node at (0.9,0) {$\ldots$};
    \node at (2.4,0) {$\ldots$};
    \node[below] at (0,-0.25) {$1$};
    \node[below] at (0.25,-0.25) {$2$};
      \node[below] at (0.5,-0.25) {$3$};
\node[below] at (1.5,-0.25) {$i$};
    \node[below] at (3,-0.25) {$N$};
\end{tikzpicture} \,.
\end{equation}

As for the TL algebra the multiplication relations between the generators follow  from their geometrical representation.
Here are a few sample relations:
\begin{eqnarray}
e_i^- e_i &=& n e_i^- \,, \nonumber \\
e_i e_i^+ &=& n e_i^+ \,, \nonumber \\
e_i^+ e_i^+ &=& 0 \,. \label{dilTLrels}
\end{eqnarray}

\subsection{The ${\rm U}(n)$ symmetry}

The ${\rm U}(n)$ symmetry (for $n \in \mathbb{N}$) of the model in the completely packed was probably first mentioned by Affleck \cite{Affleck90}. Associating particles with $n$ possible colours $a=1,\ldots,n$ on even sites, and holes with $n$ colours on odd sites, the Temperley-Lieb generators act on a pair of neighbours as $(e)_{ab}^{cd}=\delta_{ab}\delta_{cd}$, and the relation $e^2=n e$ arises simply because of the $n$ colours that can propagate along the loops. The choice of particles (resp.\ holes) on even (resp.\ odd) sites corresponds algebraically to taking the fundamental representation $\mathfrak{n}$ (resp.\ anti-fundamental $\bar{\mathfrak{n}}$) from the ${\rm U}(n)$ point of view. In the present dilute oriented case, each edge now carries  the direct sum of the trivial representation, denoted $\mathfrak{1}$ and of dimension $1$, and the fundamental (resp.\ the anti-fundamental). The space on which the transfer matrix acts is thus, for integer loop weight $n$, 
\begin{equation}
 {\cal H}=\left[(\mathfrak{1}+\mathfrak{n}) \otimes (\mathfrak{1} + \bar{\mathfrak{n}})\right]^{\otimes L}
\end{equation}
The generators can easily be written in coordinates if we associate with the representation $\mathfrak{1}$ an extra label $0$.
Introduce now Greek symbols $\alpha=0,1,\ldots,n$ to describe all the states in the trivial and fundamental representations.  We have then
\begin{eqnarray}
(e)_{\alpha\beta}^{\gamma\delta}&=&(1-\delta_{\alpha0})(1-\delta_{\beta0})
(1-\delta_{\gamma0})(1-\delta_{\delta0})\delta_{\alpha\beta}\delta_{\gamma\delta} \,, \nonumber\\
(e^+)_{\alpha\beta}^{\gamma\delta}&=&\delta_{\alpha0}\delta_{\beta0}
(1-\delta_{\gamma0})(1-\delta_{\delta0})\delta_{\gamma\delta}\nonumber \,, \\
(e^-)_{\alpha\beta}^{\gamma\delta}&=&(1-\delta_{\alpha0})(1-\delta_{\beta0})
\delta_{\gamma0}\delta_{\delta0}\delta_{\alpha\beta} \,.
\end{eqnarray}
Since the $e_i,e^\pm_i$ have non-trivial action only between the two singlets that appear in the 
tensor product $(\mathfrak{1}+\mathfrak{n})\otimes (\mathfrak{1}+\bar{\mathfrak{n}})$  (or the product with $\mathfrak{n},\bar{\mathfrak{n}}$ switched), they necessarily commute with the (trivial) action of ${\rm U}(n)$.

\subsection{Combinatorics and modules}

\subsubsection{Reminders about the completely packed case.}

In this section and the few following we will consider the case where the `space' direction for the transfer matrix  is an open segment, corresponding to open boundary conditions for the spin chain. The corresponding geometry is  a square lattice oriented diagonally with free boundary conditions in the space direction, as illustrated in figure~\ref{figdil}.

For $n= q+q^{-1}$, with $q$ generic (not a root of unity), the representation theory of the TL algebra is well known \cite{Martin_book} to be semi-simple, with all simple modules corresponding to the so-called standard modules $\widetilde{\mathcal{V}}_{j}$. These are indexed by the (even, if the number of sites $N$ is even) number $j$ of `through-lines'  (or `strings'), which are not allowed to intersect any arc.%
\footnote{Note that our convention slightly differs with the usual one in the TL litterature, where the notation $\mathcal{V}_{j}$ corresponds to the module with $2j$ through-lines.}
In both the transfer matrix and Hamiltonian pictures, the number of through-lines is semi-conserved, in the sense that it can only be lowered under the action of the TL generators (in other words, the transfer matrix and Hamiltonian have a block-triangular structure). Further imposing that pairs of through-lines cannot be contracted with one another therefore amounts to considering through-lines as conserved (or, in other words, to consider only the block-diagonal part of the transfer matrix or Hamiltonian).

\subsubsection{Standard modules of the dilute oriented case.}

Through-lines can be defined in the same way in the dilute oriented model, however they are not the only conserved quantities. Denoting the parity of edges within any row as eoeo\dots from left to right, where e stands for even, and o stands for odd, let $N_{\rm e}$
(resp.\ $N_{\rm o}$) be the number of even (resp.\ odd) edges covered by the loops within a row. Then
\begin{equation}
 Q = \frac12 \left( N_{\rm e} - N_{\rm o} \right)
\end{equation}
is conserved by the row-to-row transfer matrix. 

The standard modules ${\cal V}_j$ for this problem are not only indexed by the number $j$ of through-lines, but also by their
parity.  When, say, an even through-line is joined to an arc, it must necessarily join to the ``odd'' end of the arc (on the
nearest-neighbour site), and hence come out at the other ``even'' end of the arc. Therefore the parity of each individual through-line is
conserved by the transfer matrix. The first few standard modules are therefore labelled as follows
$${\cal V}_0 \qquad {\cal V}_1({\rm e}) \qquad {\cal V}_1({\rm o}) \qquad
{\cal V}_2({\rm eo}) \qquad {\cal V}_2({\rm ee}) \qquad {\cal V}_2({\rm oe}) \qquad {\cal V}_2({\rm oo})$$

We now compute the dimensions of these standard modules by combinatorial means.

\subsubsection{Ground state sector (no through-lines).}

Consider first the sector without through-lines, namely the standard module $\mathcal{V}_0$. It consists of
dilute arc configurations, with the crucial constraint that each arc connects two sites of opposite
parities. This implies that the number of empty sites inside any arc $A$---not counting the sites
inside other arcs $B$ contained within $A$---must be even.

Let $t$ be the weight per site, and let $f_{\rm e}(t)$ denote the generating function for valid arc
configurations on an even number of sites. We similarly define $f_{\rm o}(t)$ as the generating
function for arc configurations on a odd number of sites. By examining the possibilities for the
leftmost site we obtain the functional equations
\begin{eqnarray}
 f_{\rm e}(t) &=& 1 + t f_{\rm o}(t) + t^2 f_{\rm e}(t)^2 \,, \label{fe-genf} \\
 f_{\rm o}(t) &=& t f_{\rm e}(t) + t^2 f_{\rm e}(t) f_{\rm o}(t) \,.
\end{eqnarray}
For instance, the three terms on the right-hand side of (\ref{fe-genf}) correspond to the leftmost site being non-existent, empty,
or supporting an arc. In the latter case, the arc connects an even and an odd site, so the sites inside the arc and those following
it are independent (because of the self-avoidance) and both described by $f_{\rm e}(t)$. 

The regular solution for $f_{\rm e}(t)$ reads
\begin{eqnarray}
 f_{\rm e}(t) &=& \frac{1}{6} \left( \frac{4}{t^2} - \frac{2 (-2)^{1/3}}{\omega^{1/3}} + \frac{(-2)^{2/3} \omega^{1/3}}{t^4} \right) \,, \\
 \omega &=& -2 t^6 + 27 t^8 + 3^{3/2} t^7 \sqrt{27 t^2 - 4} \,.
\end{eqnarray}
The corresponding series expansion is
\begin{eqnarray}
 f_{\rm e}(t) &=& \sum_{L=0}^\infty a_L t^{2L} \,, \\
 a_L &=& \frac{1}{L+1} {3L+1 \choose L} \,,
 \label{eq:aL}
\end{eqnarray}
meaning that $a_L$ is the dimension of the standard module $\mathcal{V}_0$ for the model defined on an even number of sites $2L$.
For instance, $a_2 = 7$ and the seven arc configurations on four sites can be written:
\begin{equation*}
\begin{tikzpicture}[scale=0.25]
 \foreach \xpos in {0,1,2,3}
  \filldraw (\xpos,0) circle(3pt);
\begin{scope}[xshift=6cm]
 \draw[red,line width=0.5mm] (2,0) arc(180:360:5mm and 10mm);
 \foreach \xpos in {0,1,2,3}
  \filldraw (\xpos,0) circle(3pt);
\end{scope}
\begin{scope}[xshift=12cm]
 \draw[red,line width=0.5mm] (1,0) arc(180:360:5mm and 10mm);
 \foreach \xpos in {0,1,2,3}
  \filldraw (\xpos,0) circle(3pt);
\end{scope}
\begin{scope}[xshift=18cm]
 \draw[red,line width=0.5mm] (0,0) arc(180:360:5mm and 10mm);
 \foreach \xpos in {0,1,2,3}
  \filldraw (\xpos,0) circle(3pt);
\end{scope}
\begin{scope}[xshift=24cm]
 \draw[red,line width=0.5mm] (0,0) arc(180:360:15mm and 10mm);
 \foreach \xpos in {0,1,2,3}
  \filldraw (\xpos,0) circle(3pt);
\end{scope}
\begin{scope}[xshift=30cm]
 \draw[red,line width=0.5mm] (0,0) arc(180:360:5mm and 10mm);
 \draw[red,line width=0.5mm] (2,0) arc(180:360:5mm and 10mm);
 \foreach \xpos in {0,1,2,3}
  \filldraw (\xpos,0) circle(3pt);
\end{scope}
\begin{scope}[xshift=36cm]
 \draw[red,line width=0.5mm] (1,0) arc(180:360:5mm and 5mm);
 \draw[red,line width=0.5mm] (0,0) arc(180:360:15mm and 10mm);
 \foreach \xpos in {0,1,2,3}
  \filldraw (\xpos,0) circle(3pt);
\end{scope}
\end{tikzpicture}
\end{equation*}
We note in particular that the two configurations
\begin{tikzpicture}[scale=0.25]
 \draw[red,line width=0.5mm] (1,0) arc(180:360:10mm and 6mm);
 \foreach \xpos in {0,1,2,3}
  \filldraw (\xpos,0) circle(3pt);
\end{tikzpicture}
and
\begin{tikzpicture}[scale=0.25]
 \draw[red,line width=0.5mm] (0,0) arc(180:360:10mm and 6mm);
 \foreach \xpos in {0,1,2,3}
  \filldraw (\xpos,0) circle(3pt);
\end{tikzpicture}
are forbidden by the orientation constraint (the arcs connect sites with the same parity).
Those states would however be included in the standard module of the usual (not oriented)
dilute TL algebra \cite{GrimmPearce93,Grimm96} with {\rm O}($n$) symmetry.

The similar result for $f_{\rm o}(t)$ leads to
\begin{eqnarray}
 f_{\rm o}(t) &=& \sum_{L=1}^\infty b_L t^{2L-1} \,, \\
 b_L &=& \frac{1}{2L+1} {3L \choose L} \,,
\end{eqnarray}
so that now $b_L$ is the dimension of $\mathcal{V}_0$ on an odd number of sites $2L-1$.
For instance, $b_2 = 12$ and the twelve arc configurations on five sites can be written:
\begin{equation*}
\begin{tikzpicture}[scale=0.25]
 \foreach \xpos in {0,1,2,3,4}
  \filldraw (\xpos,0) circle(3pt);
\begin{scope}[xshift=7cm]
 \draw[red,line width=0.5mm] (3,0) arc(180:360:5mm and 10mm);
 \foreach \xpos in {0,1,2,3,4}
  \filldraw (\xpos,0) circle(3pt);
\end{scope}
\begin{scope}[xshift=14cm]
 \draw[red,line width=0.5mm] (2,0) arc(180:360:5mm and 10mm);
 \foreach \xpos in {0,1,2,3,4}
  \filldraw (\xpos,0) circle(3pt);
\end{scope}
\begin{scope}[xshift=21cm]
 \draw[red,line width=0.5mm] (1,0) arc(180:360:5mm and 10mm);
 \foreach \xpos in {0,1,2,3,4}
  \filldraw (\xpos,0) circle(3pt);
\end{scope}
\begin{scope}[xshift=28cm]
 \draw[red,line width=0.5mm] (0,0) arc(180:360:5mm and 10mm);
 \foreach \xpos in {0,1,2,3,4}
  \filldraw (\xpos,0) circle(3pt);
\end{scope}
\begin{scope}[xshift=35cm]
 \draw[red,line width=0.5mm] (1,0) arc(180:360:15mm and 10mm);
 \foreach \xpos in {0,1,2,3,4}
  \filldraw (\xpos,0) circle(3pt);
\end{scope}
\begin{scope}[xshift=0cm,yshift=-2.5cm]
 \draw[red,line width=0.5mm] (0,0) arc(180:360:15mm and 10mm);
 \foreach \xpos in {0,1,2,3,4}
  \filldraw (\xpos,0) circle(3pt);
\end{scope}
\begin{scope}[xshift=7cm,yshift=-2.5cm]
 \draw[red,line width=0.5mm] (1,0) arc(180:360:5mm and 10mm);
 \draw[red,line width=0.5mm] (3,0) arc(180:360:5mm and 10mm);
 \foreach \xpos in {0,1,2,3,4}
  \filldraw (\xpos,0) circle(3pt);
\end{scope}
\begin{scope}[xshift=14cm,yshift=-2.5cm]
 \draw[red,line width=0.5mm] (0,0) arc(180:360:5mm and 10mm);
 \draw[red,line width=0.5mm] (3,0) arc(180:360:5mm and 10mm);
 \foreach \xpos in {0,1,2,3,4}
  \filldraw (\xpos,0) circle(3pt);
\end{scope}
\begin{scope}[xshift=21cm,yshift=-2.5cm]
 \draw[red,line width=0.5mm] (0,0) arc(180:360:5mm and 10mm);
 \draw[red,line width=0.5mm] (2,0) arc(180:360:5mm and 10mm);
 \foreach \xpos in {0,1,2,3,4}
  \filldraw (\xpos,0) circle(3pt);
\end{scope}
\begin{scope}[xshift=28cm,yshift=-2.5cm]
 \draw[red,line width=0.5mm] (2,0) arc(180:360:5mm and 5mm);
 \draw[red,line width=0.5mm] (1,0) arc(180:360:15mm and 10mm);
 \foreach \xpos in {0,1,2,3,4}
  \filldraw (\xpos,0) circle(3pt);
\end{scope}
\begin{scope}[xshift=35cm,yshift=-2.5cm]
 \draw[red,line width=0.5mm] (1,0) arc(180:360:5mm and 5mm);
 \draw[red,line width=0.5mm] (0,0) arc(180:360:15mm and 10mm);
 \foreach \xpos in {0,1,2,3,4}
  \filldraw (\xpos,0) circle(3pt);
\end{scope}
\end{tikzpicture}
\end{equation*}

\subsubsection{With through-lines.}
\label{sec:state-counting}

In the above treatment of ${\cal V}_0$ we have considered systems of both even and odd length,
because we shall need both generating functions $f_{\rm e}(t)$ and $f_{\rm o}(t)$ in the following.
For the case with $j > 0$ through-lines, however, we limit the discussion to systems containing an even number of sites $2L$,
since this is what is needed to reproduce the geometry of figure~\ref{figdil}.

The generating function for the number of states in ${\cal V}_j(\cdots)$ is the product of
the following factors:
\begin{itemize}
 \item $f_{\rm e}$ for each pair of consecutive parity labels that are different;
 \item $f_{\rm o}$ for each pair of consecutive parity labels that are equal;
 \item $f_{\rm e}$ if the first parity label is even, and $f_{\rm o}$ if it is odd;
 \item $f_{\rm o}$ if the last parity label is even, and $f_{\rm e}$ if it is odd;
 \item $t^j$, where $j$ is the total number of through-lines.
\end{itemize}
For instance, ${\cal V}_4({\rm eoee})$ is associated with the generating function $f_{\rm e}^3 f_{\rm o}^2 t^4$.
In general, denote the sequence of parity labels by $(\mu_1 \mu_2 \cdots \mu_j)$, where $\mu_i = 1$ (resp.\ $\mu_i = -1$)
if the $i$'th label is even (resp.\ odd). We define
\begin{equation}
 \varepsilon = \sum_{i=1}^{j-1} \mu_i \mu_{i+1} \,.
\end{equation}
It follows from the above itemised list that the corresponding generating function is
\begin{equation}
 (f_{\rm e})^{\frac12(j-\epsilon+\mu_1-\mu_j+1)} (f_{\rm o})^{\frac12(j+\epsilon-\mu_1+\mu_j+1)} t^j \,.
\end{equation}

All cases of an even number of through-lines are therefore described by the generating function
\begin{equation}
 (f_{\rm e})^{2k+1} (f_{\rm o})^{2 \ell} t^{2k+2 \ell}
\end{equation}
for appropriate values of the integers $k$ and $\ell$. After extensive manipulations this expands as
\begin{equation}
 \sum_{L=k+2\ell}^\infty \frac{2k+3\ell+1}{L+k+\ell+1} {3L+k+1 \choose L-k-2\ell} t^{2L} \,,
\end{equation}
where we note that the coefficient of the initial $t^{2(k+2\ell)}$ is unity.
The dimension of the standard module ${\cal V}_j$ (for an even number of sites $2L$) therefore reads
\begin{equation}
 d_{k,\ell}  = \frac{2k+3\ell+1}{L+k+\ell+1} {3L+k+1 \choose L-k-2\ell} \,, \qquad \mbox{for } j = 2 k + 2 \ell \mbox{ even} \,,
 \label{dkell-even}
\end{equation}
where $k = \ell = 0$ for $j = 0$; and for any even $j \ge 2$:
\begin{eqnarray}
 k &=& \frac14 (j - \varepsilon + \mu_1 - \mu_j - 1) \,, \nonumber \\
 \ell &=& \frac14 (j + \varepsilon - \mu_1 + \mu_j + 1) \,. \label{kell-even}
\end{eqnarray}
In particular, for the case of zero through-lines ($k=\ell=0$) we recover from the general result (\ref{dkell-even})
the expression (\ref{eq:aL}) for the dimension $a_L$ of $\mathcal{V}_0$.

Similarly, all cases of an odd number of through-lines are described by the generating function
\begin{equation}
 (f_{\rm e})^{2k+1} (f_{\rm o})^{2 \ell+1} t^{2k+2 \ell+1} \,,
\end{equation}
with again $k, \ell \in \mathbb{N}$. Similar computations lead to the expansion
\begin{equation}
 t^{2(k+2\ell+1)} + \sum_{L=k+2\ell+2}^\infty \frac{4k+6\ell+5}{L-k-2\ell-1} {3L+k+1 \choose L-k-2\ell-2} t^{2L} \,,
\end{equation}
where we have singled out the first term, where otherwise the unit coefficient can be found by analytic continuation
of the general expression. 
The corresponding dimensions therefore read 
\begin{equation}
d_{k,\ell} = \begin{cases} 
   1       & \text{if } L=k+2\ell +1 \\
  \frac{4k+6\ell+5}{L-k-2\ell-1} {3L+k+1 \choose L-k-2\ell-2}  & \text{if } L\geq k+2\ell +2\\
  \end{cases}
  \qquad \mbox{for } j = 2 k + 2 \ell + 1 \mbox{ odd} \,,
\end{equation}
where now (\ref{kell-even}) must be replaced, for any odd $j \ge 1$, by
\begin{eqnarray}
 k &=& \frac14 (j - \varepsilon + \mu_1 - \mu_j - 1) \,, \nonumber \\
 \ell &=& \frac14 (j + \varepsilon - \mu_1 + \mu_j - 1) \,.
\end{eqnarray}

\subsection{The centraliser}

We give two different constructions of the centraliser and compute its dimension within each standard module.
The first construction is combinatorial and based on the resolution of the Markov trace in terms of traces over
standard modules. The second construction is algebraic and based on oscillator representations of the generators.

\subsubsection{Markov trace construction.}

In the usual dilute TL case, where the lattice orientation is not respected, one can find the degeneracy $D_j$ of each standard module ${\cal V}_j$ by inversion
of the sum rule \cite{JS-combi}
\begin{equation}
 \sum_{j=0}^{2L} d_j D_j = (\ell+1)^{2L} \,,
 \label{sumrule}
\end{equation}
where $d_j$ is the dimension of ${\cal V}_j$, and $\ell$ the weight of a non-contractible loop. We have $(n+1)$ states on each site,
since it can be occupied by any of the $n$ loop colours, or be empty. We shall denote by $n$ (resp.\ $\ell$) the weight of a contractible
(resp.\ non-contractible) loop. It can be seen that $D_j$ is a polynomial of degree $j$ in $\ell$,
\begin{equation}
 D_j = U_j(\ell/2) \,,
\end{equation}
where $U_j(x)$ is the $j$'th order Chebyshev polynomial of the second kind. Note in particular that $D_j$ does not depend on $n$.

The sum rule (\ref{sumrule}) is insufficient for the oriented dilute TL model investigated in this paper, since the number of sectors (standard modules)
grows exponentially, not linearly, with the size $2L$ of the system. However, the sum rule is just a special case of the more
general decomposition of the Markov trace of any element $w$ of the algebra:
\begin{equation}
 {\rm Mtr}\, w = \sum_{{\cal V}_j(\cdots)} D_j(\cdots) \, {\rm tr}_{{\cal V}_j(\cdots)} w \,,
 \label{Markovtrace}
\end{equation}
where $(\cdots)$ stands for the sequence of $j$ parity labels, and ${\rm tr}$ denotes the standard (matrix) trace.

We can now find all the $D_j(\cdots)$ by considering a suitable number of different $w$. It suffices to take words
$w$ such that a given subset of the $2L$ points are linked (by the identity operator) to the points immediately above them,
while the remaining points are constrained to be empty. In the notation of (\ref{id-strand}) we have for instance
$w =$
\begin{tikzpicture}[scale=0.7,baseline={([yshift=-.5ex]current bounding box.center)}]
 \foreach \xpos in {0,0.3,0.6}
 \draw[red,dashed,line width=0.5mm] (\xpos,-0.25) -- (\xpos,0.25);
\end{tikzpicture}
$\cdots$
\begin{tikzpicture}[scale=0.7,baseline={([yshift=-.5ex]current bounding box.center)}]
 \foreach \xpos in {0,0.3,0.6}
 \draw[black,line width=0.2mm] (\xpos,-0.25) -- (\xpos,0.25);
\end{tikzpicture}
, where the leftmost points shown are linked while the rightmost ones are empty.
One starts with the word in which all points are empty, then the word consisting of one link (residing on a
site of even or odd parity), then words with two links (of any parities), and so on.
The sum rule (\ref{sumrule}) corresponds to the particular case where all points are linked to those above them,
i.e., where $w = I$ is the identity.

The matrix traces are evaluated by placing $w$ on top of any basis element $b \in {\cal V}_j(\cdots)$ and checking if we
get back the same basis element, and if so with which weight \cite{Richard06}. Such a basis element $b$ can be said to
be `compatible' with $w$ and hence contributes to ${\rm tr}_{{\cal V}_j(\cdots)}$ with the corresponding weight.
Since through-lines have a conserved parity, we can give
two distinct weights, $\ell_{\rm e}$ and $\ell_{\rm o}$, to even and odd non contractible loops. On the left-hand side
of (\ref{Markovtrace}) we get
\begin{equation}
 (\ell_{\rm e}+1)^{N_{\rm e}} (\ell_{\rm o}+1)^{N_{\rm o}} \,,
\end{equation}
where $N_{\rm e}$ (resp. $N_{\rm o}$) is the number of even (resp.\ odd) links in $w$, i.e., the numbers of non-contractible
loops of each parity in ${\rm Mtr}\, w$.
 As explained earlier, different through-lines are not allowed to be contracted by arcs within the standard module,
and the weight of any (contractible) loops appearing in the product is $n$. In fact, with the above choices of $w$ such loops
actually cannot appear, so all weights are $0$ or $1$.

With the above choices of $w$, it is also easy to see that the basis elements contributing to ${\rm tr}_{{\cal V}_j(\cdots)} w$
are such that all the non-linked points are empty. The linked points must carry the number of through-lines with the parity labels
specified by ${\cal V}_j(\cdots)$. Any linked point not carrying a through-line can be either empty or carry an arc. Any basis
element satisfying these criteria contributes $1$ to the trace, so it suffices to count the connectivity states with these specifications.
To this end, it is very useful that we already know from the computation of the dimensions $d_j(\cdots) = d_{k,\ell}$ how to handle
the state counting (see section~\ref{sec:state-counting}).

We stress that in these computation all that matters is the appropriate connectivity states drawn on the $j$ linked sites.
In particular, the total number of points $2L$ is immaterial, and the results for $D_j(\cdots)$ are therefore valid for any $2L \ge j$.

Let us detail one sample computation determining $D_3({\rm eeo})$. We choose any $w$ where sites with the chosen labels ${\rm eeo}$
(e.g., the first, third and fourth sites) are linked to those above them, and all remaining sites are empty.
Thus
$w =$
\begin{tikzpicture}[scale=0.7,baseline={([yshift=-.5ex]current bounding box.center)}]
 \foreach \xpos in {0,0.3,0.9}
 \draw[red,dashed,line width=0.5mm] (\xpos,-0.25) -- (\xpos,0.25);
 \foreach \xpos in {0.6}
 \draw[black,line width=0.2mm] (\xpos,-0.25) -- (\xpos,0.25);
\end{tikzpicture}
$\cdots$,
but it works equally well to conduct the computation on just three sites, taking 
$w =$
\begin{tikzpicture}[scale=0.7,baseline={([yshift=-.5ex]current bounding box.center)}]
 \foreach \xpos in {0,0.3,0.6}
 \draw[red,dashed,line width=0.5mm] (\xpos,-0.25) -- (\xpos,0.25);
\end{tikzpicture}
, provided we take into account that the three links have the specified parities (eeo).
We have then
${\rm tr}_{{\cal V}_0} w = 3$ (corresponding to the states 
\begin{tikzpicture}[scale=0.25]
 \draw[white,line width=0.5mm] (0,0) arc(180:360:5mm and 8mm);
 \foreach \xpos in {0,1,2}
  \filldraw (\xpos,0) circle(3pt);
\end{tikzpicture},
\begin{tikzpicture}[scale=0.25]
 \draw[red,line width=0.5mm] (0,0) arc(180:360:10mm and 8mm);
 \foreach \xpos in {0,1,2}
  \filldraw (\xpos,0) circle(3pt);
\end{tikzpicture} or
\begin{tikzpicture}[scale=0.25]
 \draw[red,line width=0.5mm] (1,0) arc(180:360:5mm and 8mm);
 \foreach \xpos in {0,1,2}
  \filldraw (\xpos,0) circle(3pt);
\end{tikzpicture}
%
living on the linked points),
${\rm tr}_{{\cal V}_1({\rm e})} w = 3$ (corresponding to 
\begin{tikzpicture}[scale=0.25]
 \draw[red,line width=0.5mm] (0,0) -- (0,-0.8);
 \foreach \xpos in {0,1,2}
  \filldraw (\xpos,0) circle(3pt);
\end{tikzpicture},
\begin{tikzpicture}[scale=0.25]
 \draw[red,line width=0.5mm] (1,0) -- (1,-0.8);
 \foreach \xpos in {0,1,2}
  \filldraw (\xpos,0) circle(3pt);
\end{tikzpicture} or
\begin{tikzpicture}[scale=0.25]
 \draw[red,line width=0.5mm] (0,0) -- (0,-0.8);
 \draw[red,line width=0.5mm] (1,0) arc(180:360:5mm and 8mm);
 \foreach \xpos in {0,1,2}
  \filldraw (\xpos,0) circle(3pt);
\end{tikzpicture}
%
on the linked points),
${\rm tr}_{{\cal V}_1({\rm o})} w = 1$ (the only possibility being
\begin{tikzpicture}[scale=0.25]
 \draw[red,line width=0.5mm] (2,0) -- (2,-0.8);
 \foreach \xpos in {0,1,2}
  \filldraw (\xpos,0) circle(3pt);
\end{tikzpicture}),
%
${\rm tr}_{{\cal V}_2({\rm eo})} w = 2$ (corresponding to
\begin{tikzpicture}[scale=0.25]
 \draw[red,line width=0.5mm] (0,0) -- (0,-0.8);
 \draw[red,line width=0.5mm] (2,0) -- (2,-0.8);
 \foreach \xpos in {0,1,2}
  \filldraw (\xpos,0) circle(3pt);
\end{tikzpicture} or
\begin{tikzpicture}[scale=0.25]
 \draw[red,line width=0.5mm] (1,0) -- (1,-0.8);
 \draw[red,line width=0.5mm] (2,0) -- (2,-0.8);
 \foreach \xpos in {0,1,2}
  \filldraw (\xpos,0) circle(3pt);
\end{tikzpicture}),
and similarly ${\rm tr}_{{\cal V}_2({\rm ee})} w = {\rm tr}_{{\cal V}_3({\rm eeo})} w = 1$. All other traces are zero.
We have thus from (\ref{Markovtrace})
\begin{equation}
 (\ell_{\rm e}+1)^2 (\ell_{\rm o}+1) =
 3 D_0 + 3 D_1({\rm e}) + D_1({\rm o}) + 2 D_2({\rm eo}) + D_2({\rm ee}) + D_3({\rm eeo}) \,,
\end{equation}
and since all $D_j(\cdots)$ with $j < 3$ can be found from computations with fewer linked points, this eventually determines $D_3({\rm eeo})$.

In this way we find
\begin{eqnarray}
 D_0 &=& 1 \,, \nonumber \\
 D_1({\rm e}) &=& \ell_{\rm e} \,, \nonumber \\
 D_1({\rm o}) &=& \ell_{\rm o} \,, \nonumber \\
 D_2({\rm eo}) = D_2({\rm oe}) &=& \ell_{\rm e} \ell_{\rm o} - 1 \,, \nonumber \\
 D_2({\rm ee}) &=& D_1({\rm e})^2 \,, \nonumber \\
 D_2({\rm oo}) &=& D_1({\rm o})^2 \,, \nonumber \\
 D_3({\rm eoe}) &=& \ell_{\rm e}^2 \ell_{\rm o} - 2 \ell_{\rm e} \,, \nonumber \\
 D_3({\rm oeo}) &=& \ell_{\rm e} \ell_{\rm o}^2 - 2 \ell_{\rm o} \,, \nonumber \\
 D_3({\rm eoo}) = D_3({\rm ooe}) &=& D_2({\rm eo}) D_1({\rm o}) \,, \nonumber \\
 D_3({\rm eeo}) = D_3({\rm oee}) &=& D_2({\rm eo}) D_1({\rm e}) \,, \nonumber \\
 D_3({\rm eee}) &=& D_1({\rm e})^3 \,, \nonumber \\
 D_3({\rm ooo}) &=& D_1({\rm o})^3 \,, \nonumber \\
 D_4({\rm eoeo}) = D_4({\rm oeoe}) &=& \ell_{\rm e}^2 \ell_{\rm o}^2 - 3 \ell_{\rm e} \ell_{\rm o} + 1 \,, \nonumber \\
 D_4({\rm eooe}) = D_2({\rm eo})^2 \,. \label{Deo_dims}
\end{eqnarray}
From these examples the general result can now be inferred. Whenever the pattern of parity labels is alternating (eoeo\dots, or oeoe\dots)
the result is that of the usual dilute TL model, with the obvious replacements $\ell \to \ell_{\rm e}$ or $\ell_{\rm o}$. More precisely, for even $j$,
each monomial has an equal number of $\ell_{\rm e}$ and $\ell_{\rm o}$ factors, whereas for odd $j$ each monomial has one excess factor
of $\ell_{\rm e}$ (resp.\ $\ell_{\rm o}$) for the alternating pattern starting and ending with an `e' (resp. `o').

Whenever the pattern of parity labels is not alternating, there is a factorisation onto contiguous alternating sub-patterns. For example
we will have
\begin{equation}
 D_6({\rm eoeeoo}) = D_3({\rm eoe}) D_2({\rm eo}) D_1({\rm o}) \,. \label{D6-factor}
\end{equation}
This factorisation is obviously unique (assuming, of course, that each sub-pattern is of maximal length).

We stated in the beginning of this section that the $D_j(\cdots)$ cannot be inferred by applying the sumrule (\ref{sumrule}) alone.
However, now that the result has been worked out, it is a non-trivial check of the expressions for both $d_j(\cdots)$ and $D_j(\cdots)$ to
verify that the sumrule is indeed satisfied. In the case of the dilute oriented model, the sum on the left-hand side should carry
over both $j$ and the sector labels. For a given number of sites $2L$ we should only sum over sector labels that can be realised
from sub-sequences of the basic pattern ${\rm eoeo}\cdots$. Using (\ref{dkell-even}) we find, for instance when $L=2$,
\begin{eqnarray}
 & & 7 D_0({\rm e}) + 5 D_1({\rm e}) + 5 D_1({\rm o}) + 6 D_2({\rm eo}) + D_2({\rm oe}) + D_2({\rm ee}) + D_2({\rm oo}) + \nonumber \\
 & & D_3({\rm eoe}) + D_3({\rm oeo}) + D_3({\rm eeo}) + D_3({\rm eoo}) + D_4({\rm eoeo}) = (\ell_{\rm e}+1)^2 (\ell_{\rm o}+1)^2 \,, \label{dilsumrule-ex}
\end{eqnarray}
where (\ref{Deo_dims}) has been used in the last equality. Note in particular that a term like $D_3({\rm ooe})$ does not appear, because
it cannot be formed as a sub-sequence of `eoeo'.

\subsubsection{Algebraic construction.}

We consider the  system with free (open) boundary conditions, and label the sites 
$i=0,\ldots,2L-1$. Recall that we associate  an $(n+1)$-dimensional complex vector
space $V_i$, which is the direct sum $\mathfrak{1} \oplus \mathfrak{n}$ of the trivial and the fundamental for even $i$
(resp.\ trivial and dual fundamental, $\mathfrak{1} \oplus \bar{\mathfrak{n}}$, for odd $i$) representations of ${\rm U}(n)$. It is convenient to use an oscillator representation, and introduce 
 the 
operators $b_i^a$, $b_{ia}^\dagger$ for $i$ even,
$\overline{b}_{ia}$, $\overline{b}_i^{a\dagger}$ for $i$ odd, with
commutation relations $[b_i^a,b_{jb}^\dagger]=
\delta_{ij}\delta_b^a$ (with $a,b=1, \ldots, n$), and similarly
for $i$ odd.

The annihilation operators $b_i^a$,
$\overline{b}_{ia}$ destroy the singlet state (which can be considered as the oscillator vacuum), the daggers indicate
the adjoint, and the spaces $V_i$ are defined by the constraints
\begin{eqnarray}
b_{ia}^\dagger b_i^a&=& 0 \hbox{ or } 1 \quad (i\hbox{ even}),\label{constr1} \nonumber \\
\overline{b}_i^{a\dagger} \overline{b}_{ia} &=& 0\hbox{ or 1} \quad (i\hbox{
odd})\label{constr2}
\end{eqnarray}
of having at most one boson per site (we use the summation convention for
repeated indices of the same type as $a$). We define the
generators of U($n$) (or in fact of the Lie algebra gl$_n$) acting in the spaces
$V_i$ by $J_{ia}^b=b_{ia}^\dagger b_i^b$ for $i$ even,
$J_{ia}^b=-\overline{b}_i^{b\dagger} \overline{b}_{ia}$ for $i$
odd, and the commutation relations among the $J_i$'s (for each $i$)
are $i$-independent. Hence the global gl$_n$ algebra, defined by
its generators $J_a^b=\sum_i J_{ia}^b$, acts in the tensor product
$V=\otimes_{i=0}^{2L-1} V_i$.
Note that the ${\rm U}(1)$ subalgebra of gl$_n$ generated by
$J_a^a$ does not act trivially on the chain (as it counts the number of bosons), in contrast with the completely packed case studied in \cite{ReadSaleur07}.

The Temperley-Lieb generator is the
usual `Heisenberg coupling' of magnetism, and can be written in
terms of the bosonic generators as  %
\be e_i=\left\{\begin{array} {rl}
\overline{b}_{i+1}^{a\dagger}b_{ia}^\dagger
b_i^b \overline{b}_{i+1,b},&\hbox{ $i$ even,}\\
\overline{b}_i^{a\dagger}b_{i+1,a}^\dagger b_{i+1}^b
\overline{b}_{ib},&\hbox{ $i$ odd.}
\end{array}\right. \ee
The $e_i$'s are Hermitian, $e_i^\dagger=e_i$. Acting in the
constrained space $V$, they satisfy  the
relations \cite{TemperleyLieb71}
\bea e_i^2 &=& n e_i,\nonumber\\ e_i\, e_{i\pm 1}\, e_i &=&
e_i,\nonumber\\  e_i\,e_j&=&e_j\,e_i \qquad (j\neq i,\;i\pm 1).
\label{tlrel}
\eea %
While $e_i$  acts as $e_i=0$ whenever the number of bosons on one of the two neighbouring spaces is zero,  we can also introduce the generators 
\be e_i^-=\left\{\begin{array} {rl}
b_i^b \overline{b}_{i+1,b},&\hbox{ $i$ even,}\\
 b_{i+1}^b
\overline{b}_{ib},&\hbox{ $i$ odd.}
\end{array}\right. \ee
and 
\be e_i^+=\left\{\begin{array} {rl}
\overline{b}_{i+1}^{a\dagger}b_{ia}^\dagger
,&\hbox{ $i$ even,}\\
\overline{b}_i^{a\dagger}b_{i+1,a}^\dagger,&\hbox{ $i$ odd.}
\end{array}\right. \ee
which act as required in the constrained spaces. The $e_i^\pm$ are Hermitian conjugate, and ${\rm U}(n)$-invariant. 

To construct the commutant algebra explicitly, we introduce the
operators (for $k\leq 2L$)%
\be%
\mathcal{J}^{a_1a_2\ldots a_k}_{b_1b_2\ldots b_k}=\sum_{0\leq
i_1<i_2<\cdots <i_k\leq 2L-1} J_{i_1 b_1}^{a_1}J_{i_2
b_2}^{a_2}\cdots J_{i_k b_k}^{a_k} \ee%
(for $k=0$, we define $\mathcal{J}=1$, and for $k=1$,
$\mathcal{J}^a_b=J^a_b$ as defined earlier). 
We moreover  impose linear conditions, that the
contraction of one of the indices $a$ with a {\em neighbouring}
index $b$ [i.e.\ of $a_l$ with $b_{l+ 1}$ (resp., $b_{l-1}$), for
$l=1$, $2$, \ldots, $k- 1$ (resp., $l=2$, \ldots, $k$)] is zero.
This gives us a basis set $J^{a_1\ldots a_k}_{b_1\ldots b_k}$,
that is `traceless' in this sense.
For example, for $k=2$, we have %
\be%
J^{a_1 a_2}_{b_1 b_2}=\mathcal{J}^{a_1 a_2}_{b_1 b_2}%
-\frac{1}{n}\mathcal{J}^{a a_2}_{b_1 a}\delta^{a_1}_{b_2}%
-\frac{1}{n}\mathcal{J}^{a_1b}_{b b_2} \delta^{a_2}_{b_1}%
+\frac{1}{n^2}\mathcal{J}^{a b}_{b a}\delta^{a_1}_{b_2}\delta^{a_2}_{b_1}%
\ee%
and there are $(n^2-1)^2$ independent such operators. In general, there are
$(D_{k})^2$ of them, where
\be%
D_j=[j+1]_q,\label{dj}\ee 
and where
$[n]_q=q^{n-1}+q^{n-3}+\ldots+q^{-n+1}=(q^n-q^{-n})/(q-q^{-1})$ is
the $q$-deformation of any integer $n$.

The exact forms are 
\be%
J^{a_1a_2\ldots a_k}_{b_1b_2\ldots b_k}=(P^\bullet P_\bullet
\mathcal{J})^{a_1a_2\ldots a_k}_{b_1b_2\ldots b_k},\label{commdef}\ee%
where $P^\bullet$ (resp.\ $P_\bullet$) is the (Jones-Wenzl) projection
operator to the `traceless' sector on the vector space indexed
by $(a_1,b_2,\ldots)$ [resp., $(b_1,a_2,\ldots,)$], which can be
constructed recursively using the TL$_k(q)$ algebra in these
spaces (see \cite{Freedman04} for a review).

In the ordinary (completely packed) case, it is known that the $\widetilde{J}^{a_1a_2\ldots
a_k}_{b_1b_2\ldots b_k}$ generate the centraliser of the TL algebra in the sector with $k$ through-lines. The total dimension of the centraliser $\widetilde{\cA}_n(2L)$ is then
\begin{equation}
 {\rm dim} \, \widetilde{\cA}_n(2L) = \sum_{j=0 }^{L} (\widetilde{D}_{2j})^2 \,. \label{dim-commut}
\end{equation}
Note that dimensions $\widetilde{D}_{2j}$ are independent of $L$,  so the total dimension of the centraliser is determined by the upper bound in the sum, $j_{\rm max}=L$. The centraliser is simple for $n$ an integer $n\geq 2$, and the $\widetilde{D}_{2j}$ are the dimensions of its irreducible representations.

The centraliser in the dilute oriented case is then obtained by acting with the generators $J^{a_1a_2\ldots a_k}_{b_1b_2\ldots b_k}$ on the sets of through lines {\sl that can be contracted by the algebra}. This set of course depends not only on the number of through lines, but now also on the associated pattern, since only lines on odd and even sites can be contracted. For instance, for the sector `oeoe' the associated representation of the centraliser has dimension $D_{4} \equiv D_4({\rm oeoe})$, while for `oeeo' it has dimension $D_{2}^2$ and for `oeee' it has dimension $D_{2} D_{1}^2$. This factorisation phenomenon has already been discussed above; see (\ref{D6-factor}) in particular. In general the dimension is a product of $D_{k_i}$ with $\sum k_i=j$, the total number of through lines. We will denote the centraliser in this case by $\cA_n(2L)$.
Its dimension is given by an expression analogous to (\ref{dim-commut}), except that the sum has to carry over both $j$ and the associated sector labels that
arise as sub-sequences of the basic pattern ${\rm eoeo}\cdots$. For instance we find, in analogy with (\ref{dilsumrule-ex}), that for $L=2$
\begin{eqnarray}
 {\rm dim} \, \cA_n(4) &=& D_0^2 + 2 D_1^2 + 2 D_2^2 + 2 (D_1 D_1)^2 + 2 D_3^2 + 2 (D_1 D_2)^2 + D_4^2 \nonumber \\
  &=& n^8 - 2 n^6 + 3 n^4 + 2 n^2 + 4 \,.
\end{eqnarray}

\subsection{The limit $(K,\tau) \to (\frac12,2)$ and the ${\rm SU}(n+1)$ point}
\label{sec:SUn+1}

Something remarkable must happen at the particular point where the parameters $K$ and $\tau$ introduced in equation (\ref{dildiags}) are $(K,\tau) = (\frac12,2)$, where the model exhibits an extended ${\rm SU}(n+1)$ symmetry. This is most easily seen starting from the completely packed model with loop fugacity $n+1$, which obviously has an ${\rm SU}(n+1)$ symmetry. Imagine now colouring each loop in two possible ways: either black (corresponding to $n$ loop components) or `transparent' (corresponding to the last component) \cite{BloteNienhuis89}. The resulting configurations have a fugacity $n$ per black loop, while the fugacity of transparent loops is trivial, so that these need not be counted in the partition function and can hence be considered invisible indeed. As a consequence of this transformation the black loops appear diluted. This diluted model coincides exactly with the model we are discussing in this paper when $K = \frac12$ and $\tau=2$ (consider twice the right-hand side of (\ref{dildiags}), bearing in mind that the first `empty' diagram corresponds now to the two different splits of the transparent loops).

In algebraic terms, what must happen is that the centraliser of the algebra,  ${\cA}_n(2L)$, must become bigger at this special value of the coupling $\tau$, and coincide then with $\widetilde{\cA}_{n+1}(2L)$. In other words, representations of ${\cA}_n(2L)$ combine to form representations of $\widetilde{\cA}_{n+1}(2L)$. This must coincide, of course, with the appearance of extra degeneracies in the transfer matrix spectrum, such that the decomposition of the Hilbert space of total dimension $(n+1)^{2L}$ can be accomplished using both the dilute oriented  TL algebra with parameter $n$ and the ordinary TL algebra with parameter $n+1$, as well as their commutants ${\cA}_n(2L)$, and $\widetilde{\cA}_{n+1}(2L)$ respectively. 

For the ordinary TL algebra, we decompose $\left[(\mathfrak{1} \oplus \mathfrak{n}) \otimes (\mathfrak{1} \oplus \bar{\mathfrak{n}})\right]^{\otimes L}$ as
\begin{equation}
 (n+1)^{2L}=\sum_{j=0}^{L} \widetilde{D}_{2j} \widetilde{d}_{2j} \,,
\end{equation}
where the $\widetilde{D}_j$ are the dimensions of the centraliser
\begin{equation}
  \widetilde{D}_j = \widetilde{[j+1]_q} \,, \qquad
  \widetilde{[2]_q}=n+1\equiv \widetilde{n} \,.
\end{equation}
We have for instance $\widetilde{D}_{1}=\widetilde{n}$, $\widetilde{D}_2=\widetilde{n}^2-1$,
$\widetilde{D}_{3}=\widetilde{n}^3-2\widetilde{n}$ and $\widetilde{D}_4=\widetilde{n}^4-3\widetilde{n}^2+1$.
For the dilute oriented TL algebra and its centraliser, we will have a different and more complicated formula, which we will analyze explicitly for $L=2$.
We first introduce the notation
\begin{equation}
 D_j=[j+1] \,, \qquad [2]=n \,,
\end{equation}
and rewrite (\ref{dilsumrule-ex}) as
\begin{eqnarray}
 (n+1)^4 &=& {d}_0 + [{2}] ({d}_1({\rm e})+{d}_1({\rm o})) + [{3}] ({d}_2({\rm eo})+{d}_2({\rm oe})) + [{2}]^2 ({d}_2({\rm ee})+{d}_2({\rm oo})) \nonumber \\
 &+& [{4}] ({d}_3({\rm oeo})+{d}_3({\rm eoe}))+ [{2}][{3}] ({d}_3({\rm ooe})+{d}_3({\rm oee})) + [{5}] {d}_4({\rm oeoe}) \,.
\end{eqnarray}
This reads explicitly 
\begin{eqnarray}
 (n+1)^4 &=& 7 + (5+5)n + (6+1)(n^2-1) + (1+1) n^2 \nonumber\\
 &+& (1+1)(n^3-2n) + (1+1)(n^3-n) + (n^4-3n^2+1) \,.
\end{eqnarray}
For the ordinary TL we would have instead
\begin{equation}
 (n+1)^4 = \widetilde{n}^4 = 2 + 3 (\widetilde{n}^2-1) + (\widetilde{n}^4-3\widetilde{n}^2+1) \,.
\end{equation}
For this to be possible, it is necessary that eigenvalues from the dilute oriented model acquire extra degeneracies at the
point $(K,\tau) = (\frac12,2)$ so that the dimensions of the commutant become larger. Still in the case $L=2$ we find that
\begin{eqnarray}
 3 (\widetilde{n}^2-1) &=& 3 + 6 n + 3 (n^2-1) \nonumber \\
  \widetilde{n}^4-3\widetilde{n}^2+1 &=& 2 + 4 n + 4 (n^2-1) + 2 n^2 + 2 (n^3-2n) + \nonumber \\
  & & 2 (n^3-n) + (n^4-3n^2+1) \,,
\end{eqnarray}
so we see, for instance, that in the representation of dimension $d_0 = 7$ of the constrained TL algebra,
2 eigenvalues will remain non degenerate, 3 will become degenerate with some of those in the sector  with one and two non contractible legs,
and 2 will become degenerate with some of those in all the other sectors. 
This is indeed what we check numerically for generic values of $n$, and can be summed up as follows: 

\medskip
\begin{tabular}{lcrccccccc}
 & & & & \underline{$(\widetilde{d}_0 = 2) \times$} & $+$ & \underline{$(\widetilde{d}_2 = 3) \times$} & $+$ & \underline{$(\widetilde{d}_4 = 1) \times$} & \ \\
 ${d}_0$ & $=$ & 7 & $=$ & 1 & \hskip-1.1cm $\big|$ & 1 & \hskip-1.1cm $\big|$ & 2 & \hskip-0.8cm $\big|$ \\
 ${d}_1({\rm e})+{d}_1({\rm o})$ & $=$ & 10 & $=$ & & & 2 & \hskip-1.1cm $\big|$ & 4 & \hskip-0.8cm $\big|$ \\
 ${d}_2({\rm eo})+{d}_2({\rm oe})$ & $=$ & 7 & $=$ & & & 1 & \hskip-1.1cm $\big|$ & 4 & \hskip-0.8cm $\big|$ \\
 ${d}_2({\rm ee})+{d}_2({\rm oo})$ & $=$ & 2 & $=$ & & & & & 2 & \hskip-0.8cm $\big|$ \\
 ${d}_3({\rm eoe})+{d}_3({\rm oeo})$ & $=$ & 2 & $=$ & & & & & 2 & \hskip-0.8cm $\big|$ \\
 ${d}_3({\rm ooe})+{d}_3({\rm eeo})$ & $=$ & 2 & $=$ & & & & & 2 & \hskip-0.8cm $\big|$ \\
 ${d}_4({\rm eoeo})$ & $=$ & 1 & $=$ & & & & & 1 & \hskip-0.8cm $\big|$ \\
\end{tabular}
\medskip

\noindent
where the numbers appearing along the vertical bars correspond to the multiplicities with which each dense module enters the decomposition of the dilute modules. 
Finding the corresponding branching rules for general $L$ is an interesting algebraic exercise, which is however beyond the scope of this paper. 

\subsection{The periodic case}
\label{sec:periodic}

A similar procedure could be repeated in the case of a lattice with periodic boundary conditions in the horizontal direction. 

As in the dense case, some care has to be taken concerning the treatment of loops wrapping around the space-like direction.
Namely, in the sector with zero through-lines, an arc between two given points
may or may not intersect the periodic boundary condition (a vertical `seam'). These two possibilities 
(which we can depict as
\begin{tikzpicture}[scale=0.25]
 \draw[red,line width=0.5mm] (0,0) arc(180:360:5mm and 6mm);
 \foreach \xpos in {0,1}
  \filldraw (\xpos,0) circle(3pt);
\end{tikzpicture}
and
\begin{tikzpicture}[scale=0.25]
 \draw[red,line width=0.5mm] (0,0) arc(0:-90:5mm and 6mm);
 \draw[red,line width=0.5mm] (1,0) arc(180:270:5mm and 6mm);
 \foreach \xpos in {0,1}
  \filldraw (\xpos,0) circle(3pt);
\end{tikzpicture}
in terms of the usual diagrams)
can be identified---formally, by taking an algebra quotient---or
considered different, leading in the latter case to additional multiplicities. In the following we will focus on the former (quotient) case.
In the sectors with $j$ through-lines, one must also attribute a definite momentum to their winding around the seam.
Specifically, each through-line picks up a phase $\varphi = \frac{2 \pi m}{j}$ (with $m=0,1,\ldots,j-1$) when it crosses
the seam towards the right (and $-\varphi$ when it crosses the seam towards the left).
Finally, one must notice that in this periodic case modules whose parity labels differ by a cyclic permutations---for instance,
the modules `eo' and `oe'---are to be identified.

We will not go into any computational detail here, but simply observe that at the ${\rm SU}(n+1)$ point $(K,\tau) = (\frac12,2)$ the spectrum in each module can similarly be decomposed in terms of that of the periodic dense model. 
For $L=2$, for instance, we find the following:

\medskip
\hskip-0.8cm
\begin{footnotesize}
\begin{tabular}{lcrccccccccccc}
 & & & & \underline{$(\tilde{d}_0 \! = \! 2) \times$} & $+ \! \! \! \! \! \! \!$ & \underline{$(\tilde{d}_2^{(0)} \! = \! 4) \times$} & $+ \! \! \! \! \! \! \! $ & \underline{$(\tilde{d}_2^{(\pi)} \! = \! 4) \times$} & $+ \! \! \! \! \! \! \!$ & \underline{$(\tilde{d}_4^{(0)} \! = \! 1) \times$} & $+ \! \! \! \! \! \! \!$ & \underline{$(\tilde{d}_4^{(\frac{\pi}{2})} \! = \! 1) \times$} & \ \\
 ${d}_0$ & $=$ & 7 & $=$ & 1 & \hskip-0.80cm $\big|$ & 1 & \hskip-0.80cm $\big|$ &  & \hskip-0.80cm $\big|$ & 1 & \hskip-0.80cm $\big|$ & & \hskip-0.85cm $\big|$ \\
 ${d}_1({\rm e})^{(0)}+{d}_1({\rm o})^{(0)}$ & $=$ & 12 & $=$ & & & 2 & \hskip-0.80cm $\big|$ &  & \hskip-0.80cm $\big|$ & 2 & \hskip-0.80cm $\big|$ & 2 & \hskip-0.85cm $\big|$ \\
 ${d}_2({\rm oe})^{(0)}$ & $=$ & 8 & $=$ & & & 1 & \hskip-0.80cm $\big|$ &  & \hskip-0.80cm $\big|$ & 2 & \hskip-0.80cm $\big|$ & 2 & \hskip-0.85cm $\big|$ \\
 ${d}_2({\rm oe})^{(\pi)}$ & $=$ & 8 & $=$ & & & & & 1 & \hskip-0.80cm $\big|$ & 2 & \hskip-0.80cm $\big|$ & 2 & \hskip-0.85cm $\big|$ \\
 ${d}_2({\rm oo})^{(0)}+{d}_2({\rm ee})^{(0)}$ & $=$ & 2 & $=$ & & & & & & & 2 & \hskip-0.80cm $\big|$ & & \hskip-0.85cm $\big|$ \\
 ${d}_2({\rm oo})^{(\pi)}+{d}_2({\rm ee})^{(\pi)}$ & $=$ & 2 & $=$ & & & & & & & & \hskip-0.80cm $\big|$ & 2 & \hskip-0.85cm $\big|$ \\
 ${d}_3({\rm oeo})^{(0)}+{d}_3({\rm eoe})^{(0)} \! \! \! \! $ & $=$ & 4 & $=$ & & & & & & & 2 & \hskip-0.80cm $\big|$ & 2 & \hskip-0.85cm $\big|$ \\
 ${d}_3({\rm oeo})^{(\frac{2\pi}{3})}+{d}_3({\rm eoe})^{(\frac{2\pi}{3})} \! \! \! \! \! \! $ & $=$ & 4 & $=$ & & & & & & & 2 & \hskip-0.80cm $\big|$ & 2 & \hskip-0.85cm $\big|$ \\
 ${d}_4({\rm eoeo})^{(0)}$ & $=$ & 1 & $=$ & & & & & & & 1 & \hskip-0.80cm $\big|$ & & \hskip-0.85cm $\big|$ \\
 ${d}_4({\rm eoeo})^{(\frac{\pi}{2})}$ & $=$ & 1 & $=$ & & & & & & & & \hskip-0.80cm $\big|$ & 1 & \hskip-0.85cm $\big|$ \\
 ${d}_4({\rm eoeo})^{(\pi)}$ & $=$ & 1 & $=$ & & & & & & & 1 & \hskip-0.80cm $\big|$ & & \\
\end{tabular}
\end{footnotesize}

\noindent
where for each module with through-lines we have indicated between parentheses the value of the corresponding phase $\varphi$, as defined above. 

\section{Low-energy limit and critical points}
\label{sec:cft}

\subsection{Mapping onto a field theory}

\label{sec:mappingfieldtheory}

In order to understand what kind of critical point one may observe in this model, it is useful to start by considering in more detail the map  of the completely packed theory onto the ${\rm CP}^{n-1}$ model.  We thus go back to (\ref{BWexp}),  introduce a continuous variable $\vec{z}(x,y)$, and expand this variable around the vertex. For instance we write 
\begin{equation}
\vec{z}_A=\vec{z}+a(-\partial_x\vec{z}+\partial_y\vec{z})+{a^2\over 2}(\partial^2_x\vec{z}+\partial^2_y\vec{z}-2\partial_x\partial_y\vec{z})+\ldots \,,
\end{equation}
where $a$ is the cutoff and we restricted to second order. A painful but straightforward calculation gives
\begin{equation}
(\vec{z}_A^\dagger \cdot \vec{z}_B)(\vec{z}_C^\dagger \cdot \vec{z}_D)=1-4a^2\left[(\vec{z} \cdot \partial_y\vec{z}^\dagger)^2+(\partial_y\vec{z}.\partial_y\vec{z}^\dagger)\right]+\ldots
\end{equation}
and similarly for the other term up to the exchange $x\leftrightarrow y$. At the isotropic point $p=1/2$ to which we will restrict, we see that the Boltzmann weight ${\rm e}^{-S}$ can be re-exponentiated, giving
\begin{equation}
S\propto  a^2\sum_{\rm vertices}(\vec{z} \cdot \partial_\mu\vec{z}^\dagger)^2+(\partial_\mu\vec{z} \cdot \partial_\mu\vec{z}^\dagger)\approx \int {\rm d}x{\rm d}y \, \left[(\vec{z} \cdot \partial_\mu\vec{z}^\dagger)^2+(\partial_\mu\vec{z} \cdot \partial_\mu\vec{z}^\dagger)\right] \,,
\end{equation}
where $\mu=x,y$ and the sum over both directions is implicit. This action can be checked to coincide with the standard action for the ${\rm CP}^{n-1}$ model \cite{ReadSaleur01}%
\footnote{The bare coupling  is of order $O(1)$ and entirely determined by the model, which has no free parameters (at the isotropic point $p={1\over 2}$). Of course, this coupling  flows under renormalisation.} 
\begin{equation}
S=\int {\rm d}x{\rm d}y \, (D_\mu\vec{z})^\dagger (D_\mu\vec{z}) \,,
\end{equation}
where the covariant derivative is 
\begin{equation}
D_\mu \equiv\partial_\mu-\vec{z}^\dagger \cdot \partial_\mu\vec{z} \,.
\end{equation}
This action is invariant under the gauge transformations $\vec{z}(x,y)\to e^{i\theta(x,y)}\vec{z}(x,y)$.

We now consider what happens for the dilute oriented model. We must add to the Boltzmann weight expansion (\ref{BWexp}) a new kind of term, which corresponds
to the possibility of having no loop or only one arch of a loop. We thus take now
\begin{eqnarray}
e^{-S}=\ldots \left\{1+\alpha\left[(\vec{z}_A^\dagger \cdot \vec{z}_B)(\vec{z}_C^\dagger \cdot \vec{z}_D)+
(\vec{z}_A^\dagger \cdot \vec{z}_D)(\vec{z}_C^\dagger \cdot \vec{z}_B)\right]\right.\nonumber\\
\left.+\beta \left[(\vec{z}_A^\dagger \cdot \vec{z}_B)+(\vec{z}_C^\dagger \cdot \vec{z}_D)+
(\vec{z}_A^\dagger \cdot \vec{z}_D)+(\vec{z}_C^\dagger \cdot \vec{z}_B)\right]\right\}\ldots
\end{eqnarray}
The same calculation as before gives now
\begin{equation}
(\vec{z}_A^\dagger \cdot \vec{z}_B)+(\vec{z}_C^\dagger \cdot \vec{z}_D)=2-4a^2\partial_x\vec{z} \cdot \partial_x\vec{z}^\dagger \,,
\end{equation}
so we see that, if we re-exponentiate in the action, we will get generally
\begin{equation}
S\propto \int {\rm d}x{\rm d}y \, \left[ \alpha(D_\mu\vec{z})^\dagger (D_\mu\vec{z})+\beta (\partial_\mu\vec{z})^\dagger (\partial_\mu\vec{z})\right] \,.
\end{equation}
This is a model with two parameters, in agreement with the fact that we have two parameters $\tau,K$ on the lattice. The important fact is that the new term $(\partial_\mu\vec{z})^\dagger (\partial_\mu\vec{z})$ {\sl breaks gauge invariance}. 

The situation we are encountering has in fact a long history in the area of sigma models, in particular in what is called `isovector-isotensor' models.
For instance, in \cite{SokalStarinets01} a model with the following lattice hamiltonian is studied
\begin{equation}
H=-\sum_{\langle xy \rangle} \left[\beta_V \vec{\sigma}_x \cdot \vec{\sigma}_y+{\beta_T\over 2}\left(
\vec{\sigma}_x \cdot \vec{\sigma}_y\right)^2\right] \,,
\end{equation}
where $\vec{\sigma}$ is an $n$-component real  spin obeying $\vec{\sigma}_x \cdot \vec{\sigma}_x=1$. The case $\beta_T=0$ is the usual `${\rm O}(n)$ model' (or, more precisely, the model with target $S^{n-1}={\rm O}(n)/{\rm O}(n-1)$), while the case $\beta_V=0$ is the ${\rm RP}^{n-1}$ model. This latter case has an extra {\sl local} $Z_2$ (gauge) symmetry, since we can change  the sign of $\vec{\sigma}$ at each point independently.%
\footnote{Note  that in the usual  ${\rm O}(n)$ model on the hexagonal lattice, once one has restricted to closed planar loops, every spin $\vec{S}$ occurs twice, and one can also change its sign at will. But this  ${\rm O}(n)$ model is not the same as the ${\rm RP}^{n-1}$ model because the action is not gauge invariant. The same occurs in our case: if we keep only the loop diagrams, the phases of the $\vec{z}$ still cancel out. But the continuous action is not gauge invariant.}

Unfortunately there are not  many results available on the phase diagram of similar deformations for the ${\rm CP}^{n-1}$ model. In the ${\rm RP}^{n-1}$ case,  it seems that   the physics of the model with both $\beta_V,\beta_T\neq 0$ is the same as that of the ordinary ${\rm O}(n)$ model \cite{SokalStarinets01}. Since ${\rm RP}^{n-1}=S^{n-1}/Z_2$, we see that  mixing of the isovector term renders the target bigger, ${\rm RP}^{n-1}\rightarrow S^{n-1}$. By analogy, we expect  that in the case of complex vectors we have 
\begin{equation}
{\rm CP}^{n-1}={\rm SU}(n)/{\rm SU}(n-1)\otimes {\rm U}(1)\rightarrow {\rm SU}(n)/{\rm SU}(n-1)=S^{2n-1} \,,
\end{equation}
that is, one should expect to observe---at least in some regions of  the phase diagram---the physics of the ${\rm O}(2n)$ model.

\subsection{Continuum limit at the ${\rm SU}(n+1)$ point}

\label{sec:SUn1:continuum}

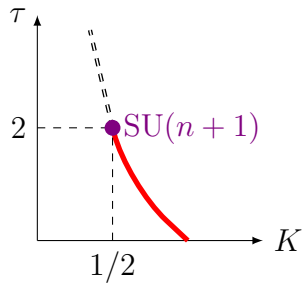
\begin{figure}
\begin{center}
\begin{tikzpicture}
\draw[->,>=latex] (0,0) -- (0,3) node[left] {$\tau$};
\draw[->,>=latex] (0,0) -- (3,0) node[right] {$K$};
\draw[red,line width=2pt,rounded corners=20pt](2,0) -- (1.2,0.75) -- (1,1.5);
\draw[double,dashed,line width=0.5pt,rounded corners=20pt] (1,1.5) -- (0.7,2.8);
\draw[dashed] (1,0) node[below] {$1/2$} -- (1,1.5);
\draw[dashed]  (0,1.5) node[left] {$2$} -- (1,1.5);
\fill[violet] (1,1.5) circle[radius=3pt] node[right] {${\rm SU}(n+1)$};
\end{tikzpicture}
\end{center}
\caption{Qualitative phase diagram of the oriented dilute loop model for some fixed value of $n \le 1$. For small monomer fugacity $K$ the system is in a massive phase with zero average density of monomer in the thermodynamic limit. For large values of $K$ one enters the so-called `dense' phase, which is critical and described by the dense ${\rm O}(n)$ universality class. The two phases are separated for low $\tau$ by a critical line (thick red curve), and for large $\tau$ by a first-order line (double dashed curve), in between which sits the ${\rm SU}(n+1)$ point (purple dot).}                                 
\label{fig:phasediag}
  \end{figure}

The phase diagram of the  dilute  oriented loop model in the $(K,\tau)$ plane for any $-1 \le n < 1$
is in fact quite simple.
Its shape is represented schematically in figure  \ref{fig:phasediag}, and it exhibits
in particular a single critical line separating a dense and a dilute phase.%
\footnote{In the following we consider that the continuous variable $n$ takes some fixed, given value,
and discuss the phase diagram in the $(K,\tau)$ plane that intersects this value of $n$ in the full $(n,K,\tau)$ space.
In particular, we shall refer to the ${\rm SU}(n+1)$ ``point'' (although it is rather a ``line'', as a function of $n$).}

Before examining this critical line in detail, we focus in this section on the ${\rm SU}(n+1)$ point $(K,\tau) = (\frac12,2)$; see section \ref{sec:SUn+1}.
It is indeed straightforward to understand the critical content of the dilute oriented model at that point  in terms of the usual  ${\rm SU}(n+1)$ dense critical model.
Let us first recall a few facts about the latter. Parameterising
\begin{equation}
\widetilde{n} = n+1 = - 2 \cos \pi \widetilde{g} \,, \qquad \mbox{with } 0 \leq \widetilde{g} \leq 1 \,,
\end{equation} 
the central charge---which is related to the finite-size scaling of the ground state (i.e., the lowest-energy eigenstate of the Hamiltonian in the sector with zero through-lines)---reads
\begin{equation}
\widetilde{c}= 1 - 6\frac{(1-\widetilde{g})^2}{\widetilde{g}} \,,
\end{equation}
and the critical exponents---which are related to the finite-size scaling of the gaps between excited eigenlevels and the ground state---can be expressed in terms of a Kac-like formula
\begin{equation}
h_{r,s} = \frac{(s \widetilde{g} - r)^2-(1-\widetilde{g})^2}{4\widetilde{g}} \,.
\end{equation}
In particular the leading exponent in the sector with $\ell =j$ through-lines (the so-called watermelon exponent $\tilde{x}_{\ell}$) reads, in the open case \cite{DuplantierSaleur86,BatchelorSuzuki93}
\begin{equation}
\tilde{x}_{\ell} = h_{1,\ell+1} =  \frac{\left[(\ell + 1) \tilde{g} - 1\right]^2-(1-\tilde{g})^2}{4\tilde{g}} \,,
\label{eq:tildexl:open}
\end{equation}
and in the periodic case \cite{DuplantierSaleur87}
\begin{equation}
\tilde{x}_{\ell} = 2 h_{0,\frac{\ell}{2}} = \frac{\ell^2 \tilde{g}}{8} - \frac{(1-\tilde{g})^2}{2\tilde{g}} \,.
\label{eq:tildexl:periodic}
\end{equation}

We now go back to the dilute oriented model. Quite generally, we can similarly define scaling exponents
$x_{\rm e}, x_{\rm o}, x_{\rm eo},\ldots$ associated with the scaling of the leading eigenvalue (lowest energy) in each sector
of $j$ through-lines with prescribed parities.%
\footnote{For simplicity, we suppress $j$ in this notation, since its value can be inferred by counting the number of parity labels.}
At the special point $(K,\tau=\frac12,2)$ these exponents should be related to the $\tilde{x}_\ell$ of the dense model.
We have checked numerically that the correspondence reads
\begin{eqnarray}
x_{\rm e} = x_{\rm o} &=& \widetilde{x}_2 \,, \nonumber\\
x_{\rm eo} = x_{\rm oe} &=& \widetilde{x}_2 \,, \nonumber\\ 
x_{\rm ee} = x_{\rm oo} &=& \widetilde{x}_{4}  \,, 
\label{eq:correspondencexeoxtilde}
\end{eqnarray} 
and so on. The general rule is that, given a sequence of parities in the dilute model, the corresponding number of through-lines in the completely packed model is obtained by filling in the minimal number of through-lines such that the sequence becomes alternating (eoeo$\cdots$eo). This is also in agreement with the decompositions in sections \ref{sec:SUn+1} and \ref{sec:periodic}

In conclusion, the exponents of the dilute oriented model at the special ${\rm SU}(n+1)$ point are the watermelon exponents $\widetilde{x}_{2j}$ of the associated completely packed model, where the (even) number $j$ of through-lines is dictated by parity constraints. 

\subsection{Numerical evidence for the ${\rm O}(2n)$ universality class}

We now turn back to the study of the critical line below the ${\rm SU}(n+1)$ point, that is,
the red line in figure \ref{fig:phasediag} for generic values $\tau <2$. 
We also shall restrict from now on to the bulk case, corresponding to a periodic Hamiltonian. In the boundary case different kinds of critical behaviour can be observed, depending on the weight given to boundary loops \cite{JacobsenSaleur08,DubailJacobsenSaleur10}.
In conjuction with the already quite complex bulk behaviour, which we shall describe below, this makes a clear convergence of the critical exponents to any of the natural candidates very difficult to assess, at least for the range of system sizes considered here
(namely, up to $2L = 18$). Accordingly, we shall leave the examination of the boundary criticality to future work.

\begin{figure}
\begin{center}
\includegraphics{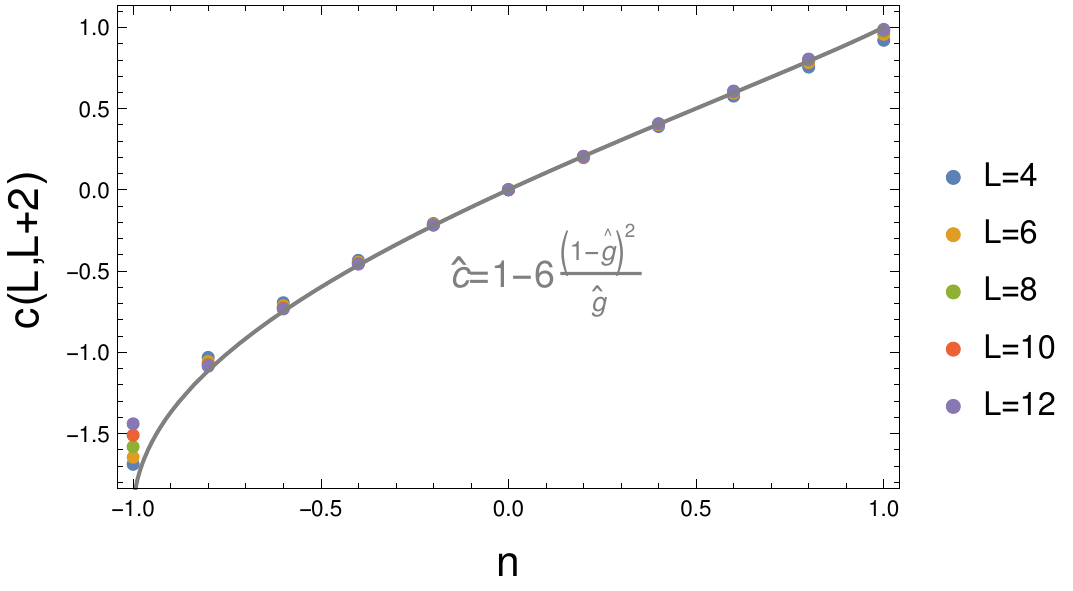}
\end{center}
\caption{
Central charge on the dilute critical line (we took here $\tau=1$), measured from finite size scaling of the transfer matrix eigenvalues at sizes $L,L+2$ for $L=4,6,8,10,12$, plotted as a function of $n$. The data shows remarkable agreement with the ${\rm O}(2n)$ central charge (\ref{eq:chat}), plotted in comparison.
}                                 
\label{fig:cO2n}
\end{figure}

Clear evidence for the ${\rm O}(2n)$ critical behaviour suggested by the analysis of section \ref{sec:mappingfieldtheory} is obtained by looking at the central charge, whose numerical estimation as a function of $n$ shows remarkable agreement with that of the ${\rm O}(2n)$ dilute branch, namely
\begin{equation}
\hat{c}=1-6\frac{(1-\hat{g})^2}{\hat{g}} \,, \qquad 2 n=-2\cos(\pi  \hat{g})\,, \qquad 1\leq \hat{g} \leq 2 \,.
\label{eq:chat}
\end{equation}
(see figure \ref{fig:cO2n}). Here and below we place a circumflex ($\hat{\ }$) on top of quantities referring to the dilute ${\rm O}(2n)$ model.
We note that  values of the  central charge for  the particular case $\tau=1$ were  already  displayed in Table V of \cite{FuGuoBlote13};
the relationship with dilute ${\rm O}(2n)$ criticality was however not noticed in this reference.

To proceed further,  we recall that the  bulk watermelon exponents  at the ${\rm O}(2n)$ critical point read
\begin{equation}
\hat{x}_{\ell} = \frac{\ell^2 \hat{g}}{8}-\frac{(1-\hat{g})^2}{2\hat{g}} \,.
\label{eq:xl:dilute}
\end{equation}
In the following sections, we will give numerical evidence for the fact that the exponents of the dilute oriented model on its critical line---that
is, on the red curve $K = K_{\rm c}(\tau)$ in figure~\ref{fig:phasediag})---are indeed given by (\ref{eq:xl:dilute}) for any value of $\tau < 2$.
Here $\ell$ simply equals the number of through-lines, regardless of their parities, and the parity constraints found in the above
discussion of the {\rm SU}($n+1$) simply disappear on the dilute critical line. This means that (\ref{eq:correspondencexeoxtilde}) is replaced by
\begin{eqnarray}
 x_{\rm e} = x_{\rm o} &=& \hat{x}_1 \nonumber\\
 x_{\rm eo} = x_{\rm oe} = x_{\rm ee} = x_{\rm oo} &=& \hat{x}_2 \,. \label{eq:correspondencexeox}
\end{eqnarray} 

\begin{figure}
\begin{center}
\begin{tikzpicture}[scale=1.5]
\fill[violet] (0,0.75) circle[radius=2pt] node[left] {$\widetilde{x}_1$};
\fill[violet] (0,1.5) circle[radius=2pt] node[left] {$\widetilde{x}_2$};
\fill[violet] (0,2.25) circle[radius=2pt] node[left] {$\widetilde{x}_3$};
\fill[violet] (0,3) circle[radius=2pt] node[left] {$\widetilde{x}_4$};
\node at (0,-0.5) {${\rm SU}(n+1)$ point};
\draw[red,line width=2pt] (4,0.5) -- (4.5,0.5) node[right] {$\hat{x}_1$};
\draw[red,line width=2pt] (4,1.2) -- (4.5,1.2) node[right] {$\hat{x}_2$};
\draw[red,line width=2pt] (4,1.9) -- (4.5,1.9) node[right] {$\hat{x}_3$};
\draw[red,line width=2pt] (4,2.6) -- (4.5,2.6) node[right] {$\hat{x}_4$};
\node at (4.5,-0.5) {dilute critical line};
\draw[red,line width=1pt,->,>=latex] (0.1,1.5) -- node[below] {$x_{\rm e}, x_{\rm o}$} (3.9,0.5);
\draw[red,line width=1pt,->,>=latex] (0.1,1.5) -- node[above] {$x_{\rm oe}, x_{\rm eo}$}  (3.9,1.2);
\draw[red,line width=1pt,->,>=latex] (0.1,3) -- node[left] {$x_{\rm ee}, x_{\rm oo}$}  (3.9,1.2);
\draw[red,line width=1pt,->,>=latex] (0.1,3) -- node[right=2.75cm,below=-0.05cm] {$x_{\rm eoe}, x_{\rm eoo}, x_{\rm eeo}, x_{\rm oeo}$}  (3.9,1.9);
\draw[red,line width=1pt,->,>=latex] (0.1,3) -- node[above] {$x_{\rm oeoe}$}  (3.9,2.6);
\end{tikzpicture}
\end{center}
\caption{Evolution of the dilute oriented model's critical exponents from the ${\rm SU}(n+1)$ point to the dilute critical line, in the case of periodic boundary conditions. The exponents $\widetilde{x}_\ell$ and $\hat{x}_\ell$ are given by (\ref{eq:tildexl:periodic}) and (\ref{eq:xl:dilute}) respectively. An analogous diagram can be drawn in the open case, with the exponents $\widetilde{x}_\ell$ and $\hat{x}_\ell$ being now given by (\ref{eq:tildexl:open}) and (\ref{eq:hatxl:open}) respectively.
}                                 
\label{fig:xxtilde}
  \end{figure}
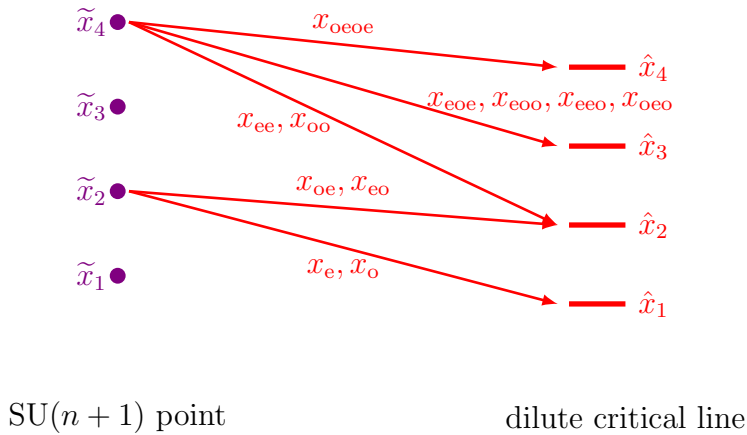

In spite of the numerical problems evoked above, we would expect---by analogy with the bulk case---that the exponents with
open boundary conditions are still those (\ref{eq:tildexl:open}) of the ordinary surface transition, which we rewrite here as
\begin{equation}
 \hat{x}_{\ell} =  \frac{\left[(\ell + 1) \hat{g} - 1\right]^2-(1-\hat{g})^2}{4\hat{g}} \,,
\label{eq:hatxl:open}
\end{equation}
with $j=\ell$ and no parity constraints. The situation is summarised in figure \ref{fig:xxtilde}.

Let us proceed to check
 the prediction (\ref{eq:correspondencexeox}) for the bulk exponents
numerically for arbitrary values of $n<1$ and $\tau<2$. For reasons that will be made clear in the following, it is convenient to first look at values of $n$ and $\tau$ far enough from $n=1$ and $\tau=2$, respectively.

\subsubsection{First look: negative $n$, small $\tau$.} 

We therefore take, for instance, $n=-\frac12$ and $\tau=0$, and compute the critical exponents by exact diagonalisation of the loop transfer matrix, by varying $K$ across the critical value $K=K_{\rm c}(\tau)$, for system sizes ranging up to $2L=16$. 
 
\begin{figure}
\begin{center}
\includegraphics[scale=0.75]{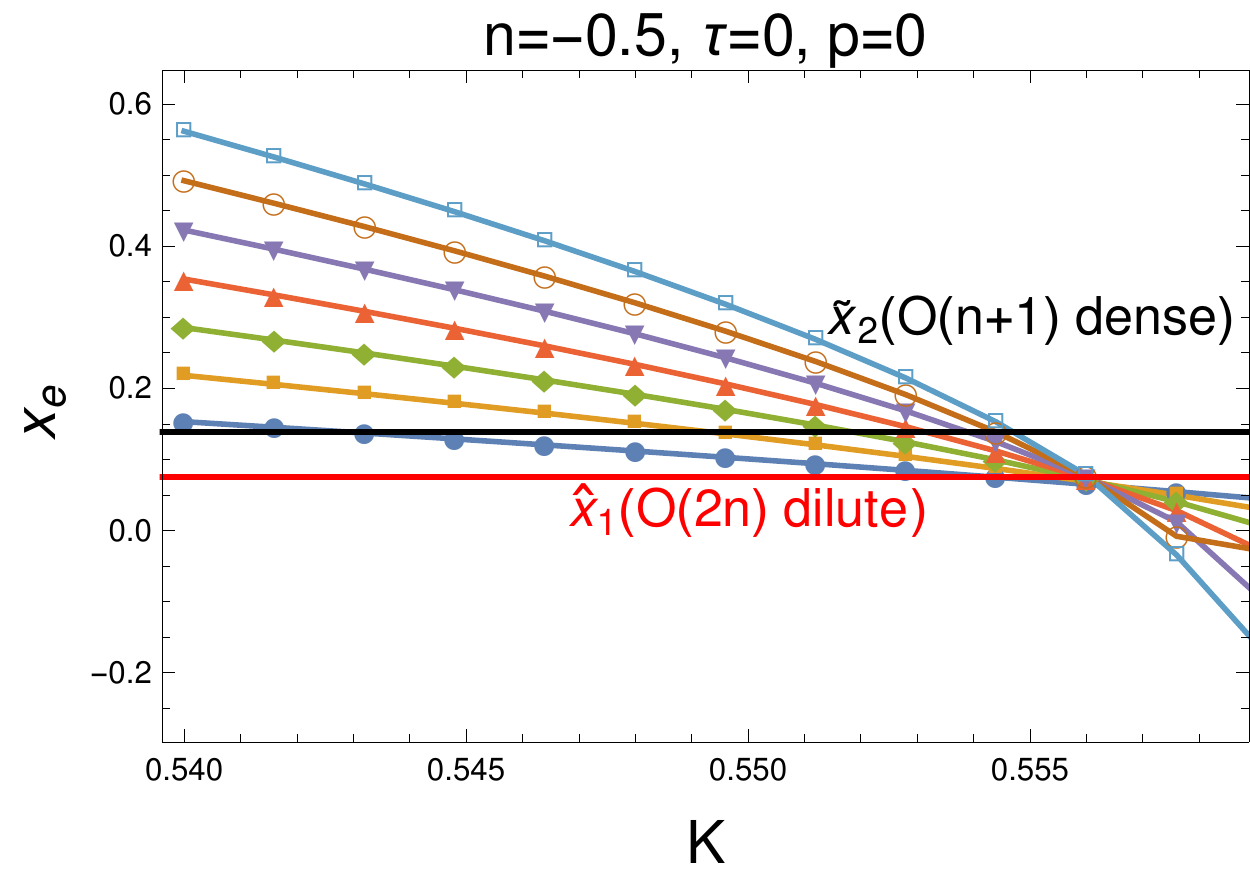}
\end{center}
\caption{Watermelon exponent $x_{\rm e} = x_{\rm o}$ across the $n=-\frac12$ dilute to dense transition, at $\tau=0$, obtained from exact diagonalisation of the periodic transfer matrix for system sizes $2L=4$ (bottom curve on the left of the graph) to $2L=16$ (top curve on the left on the graph). We show for comparison the exponent $\widetilde{x}_2$ for the dense ${\rm O}(n+1)$ model (in black), and the exponent $\hat{x}_1 = \frac{3}{40}$ for the dilute ${\rm O}(2n)$ model (in red). Convergence towards the latter is almost perfect.
}                                 
\label{fig:tau0_xe_nm05}
  \end{figure}

\begin{figure}
\begin{center}
\includegraphics[scale=0.65]{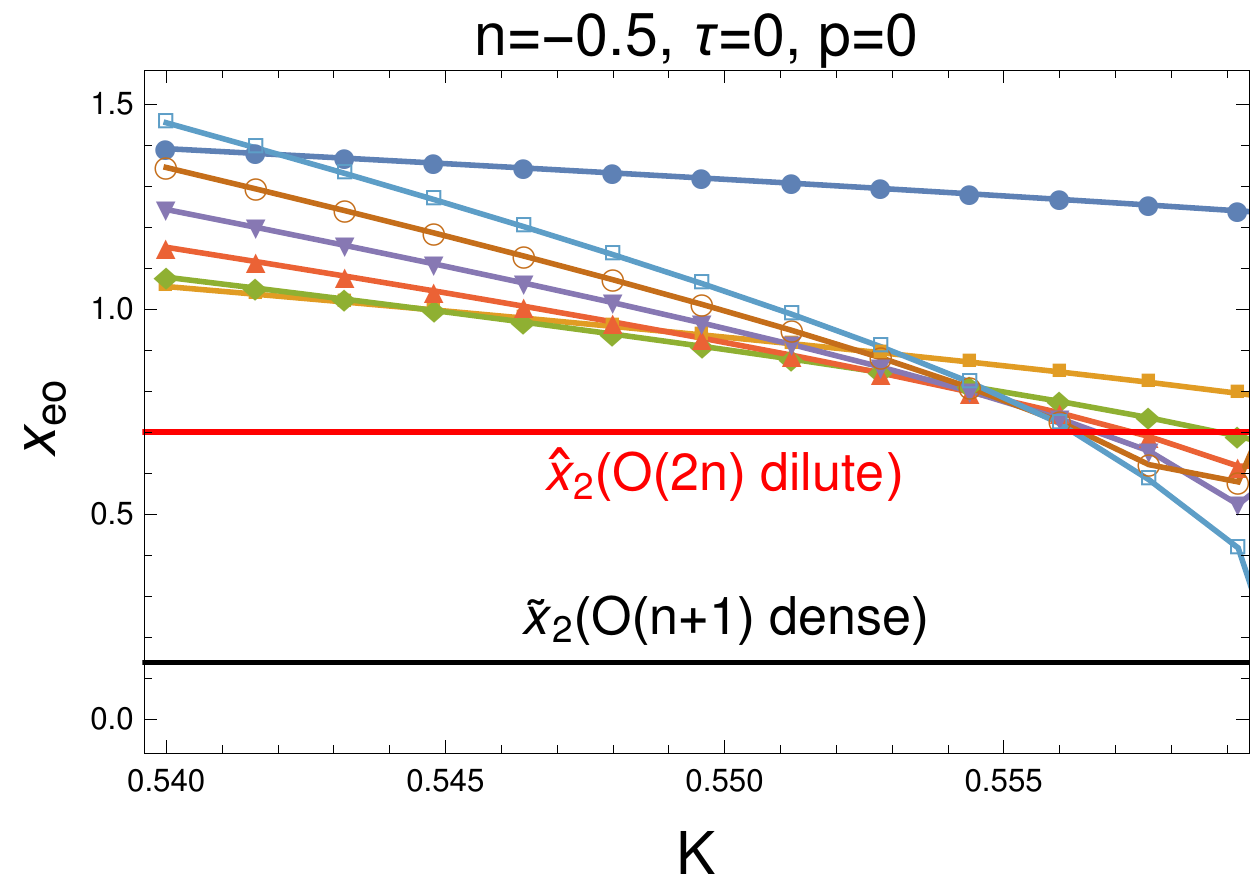}
\includegraphics[scale=0.5]{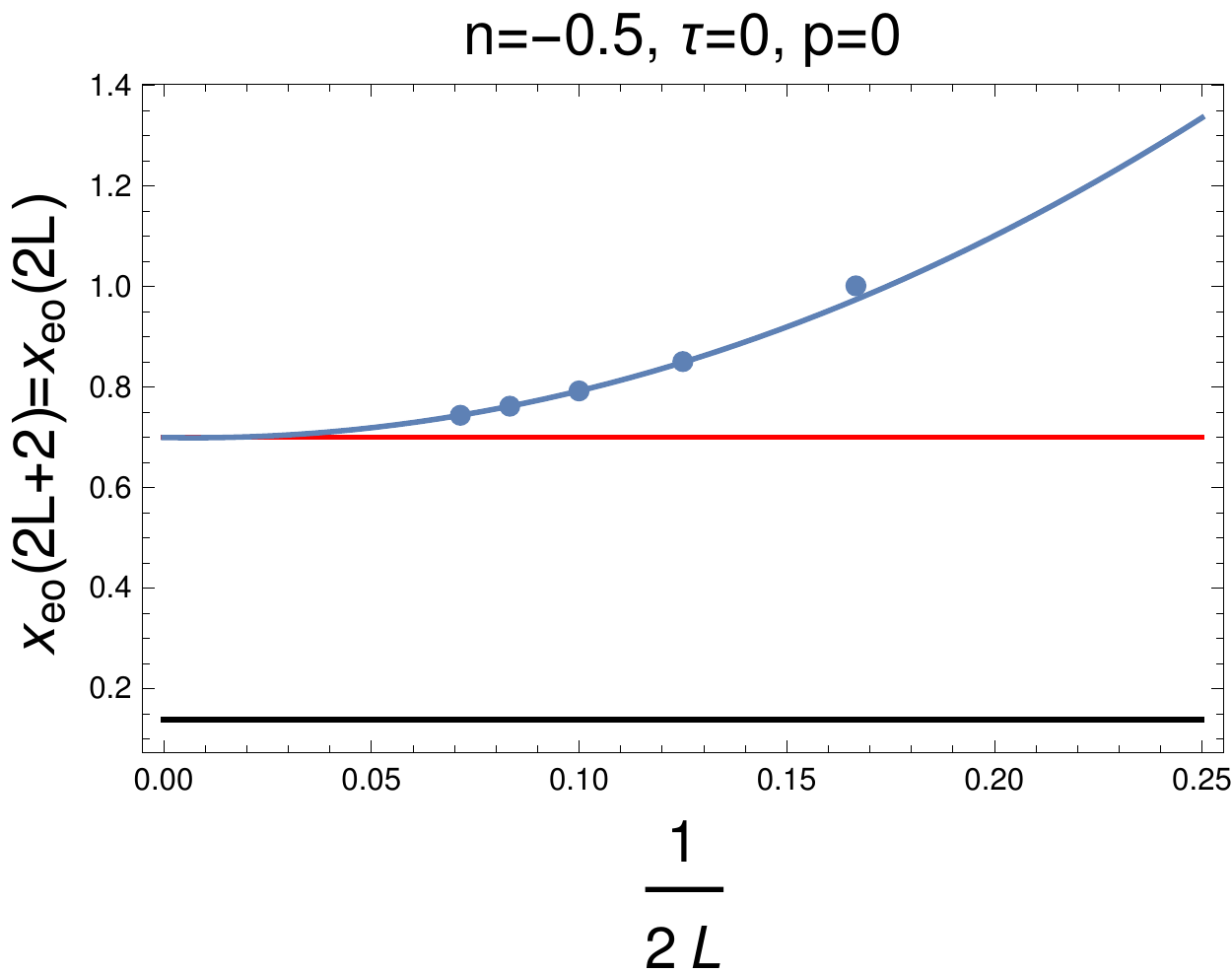}
\end{center}
\caption{Watermelon exponent $x_{\rm oe}$ across the $n=-\frac12$ dilute to dense transition, at $\tau=0$ (periodic boundary conditions). We show for comparison the exponent $\widetilde{x}_2$ for the dense ${\rm O}(n+1)$ model (in black), and the exponent $\hat{x}_2$ for the dilute ${\rm O}(2n)$ model (in red). Looking more closely at the intersections in the left figure one sees that all curves corresponding to different sizes do not intersect at the same location. So in the right panel we study the behaviour of the intersection of the exponents for sizes $2L$ and $2L+2$ as a function of $L$. The blue curves fits the last three sizes ($2L = 10,12,14$) to a parabolic function, and yields an estimate at $L \to \infty$ which coincides precisely with $\hat{x}_2 = \frac{7}{10}$ in the dilute ${\rm O}(2n)$ model.
}                                 
\label{fig:tau0_xeo_nm05}
\end{figure}

\begin{figure}
\begin{center}
\includegraphics[scale=0.65]{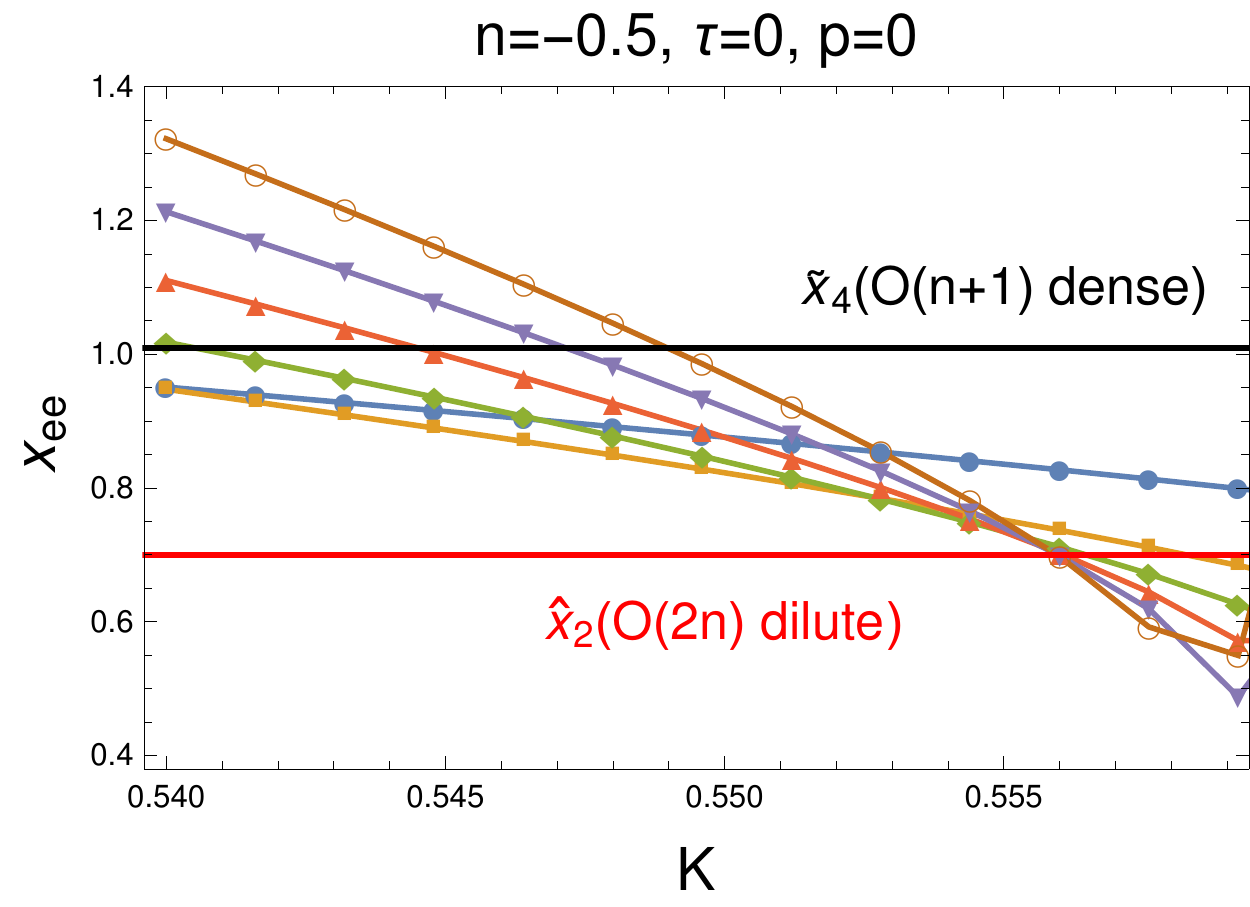}
\includegraphics[scale=0.5]{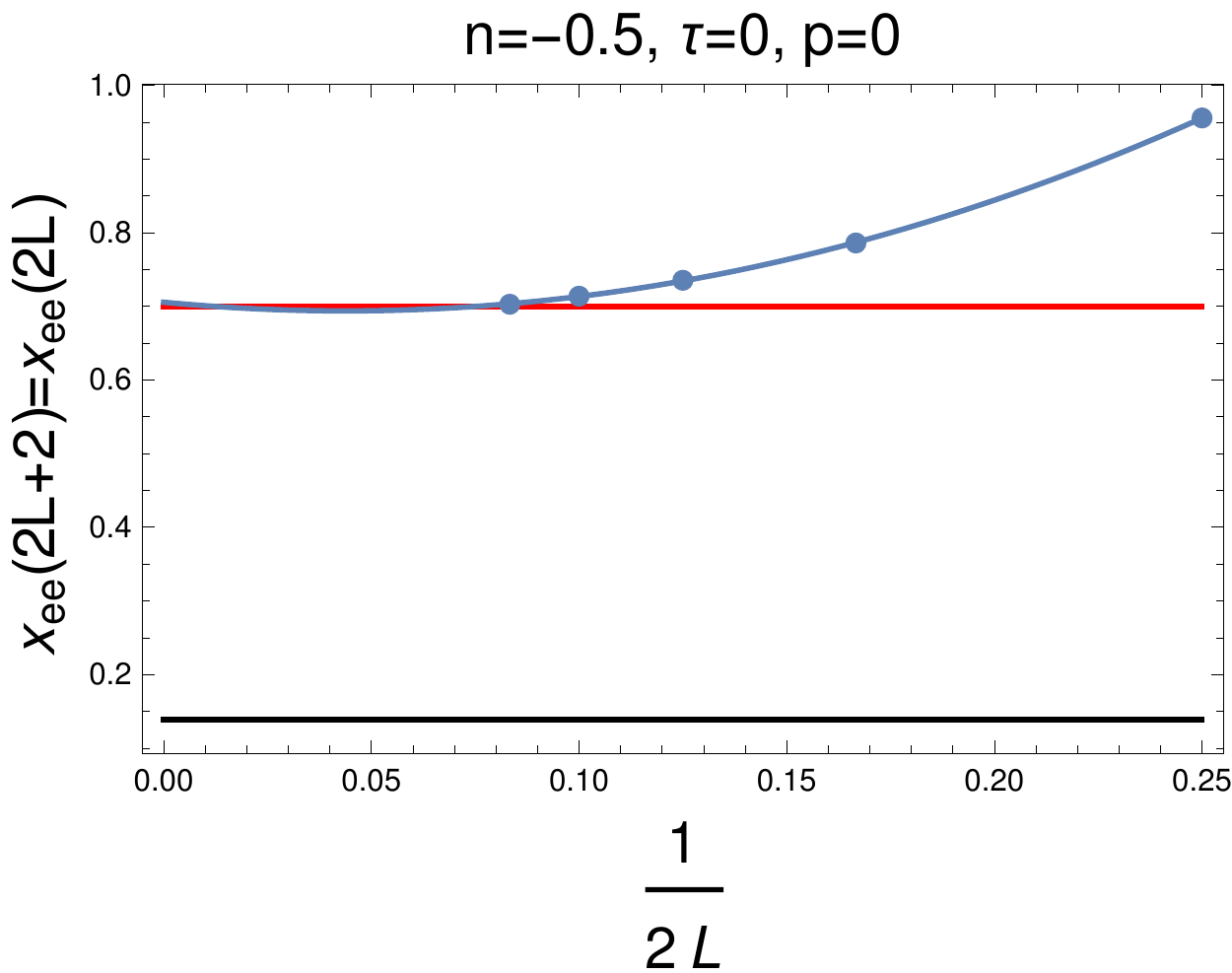}
\end{center}
\caption{The same as figure~\ref{fig:tau0_xeo_nm05}, but for the exponent $x_{\rm ee}$. The black (resp.\ red) lines shows
$\widetilde{x}_4$ for the dense {\rm O}($n+1$) model (resp. $\hat{x}_2$ for the dilute {\rm O}($2n$) model). The right panel provides the
intersection of the exponents for sizes $2L$ and $2L+2$, and the blue curves fits those values for $2L = 4,6,8,10,12$ to a parabolic function. The resulting estimate at $L \to \infty$ coincides precisely with $\hat{x}_2 = \frac{7}{10}$ in the dilute {\rm O}($2n$) model.
}                                 
\label{fig:tau0_xee_nm05}
\end{figure}

As shown in figure \ref{fig:tau0_xe_nm05}, the convergence of the exponent $x_{\rm e} = x_{\rm o}$ at the critical value $K=K_{\rm c}(\tau)$ is in perfect agreement with the {\rm O}($2n$) value $\hat{x}_1$ given by (\ref{eq:xl:dilute}). 
While for the exponents $x_{\rm ee}=x_{\rm oo}$ and $x_{\rm eo}=x_{\rm oe}$ the convergence is not as good (see figures \ref{fig:tau0_xeo_nm05} and \ref{fig:tau0_xee_nm05}), fitting the finite-size results to a quadratic function of $L^{-1}$ leaves little doubt that these exponents are once again described by (\ref{eq:correspondencexeox}).  

\subsubsection{Increasing $n$ and $\tau$.}

After these promising conclusions for negative $n$, we now look at larger values of both $n$ and $\tau$, for instance $n=\frac12$ and $\tau=1$. 
\begin{figure}
\begin{center}
\includegraphics[scale=0.65]{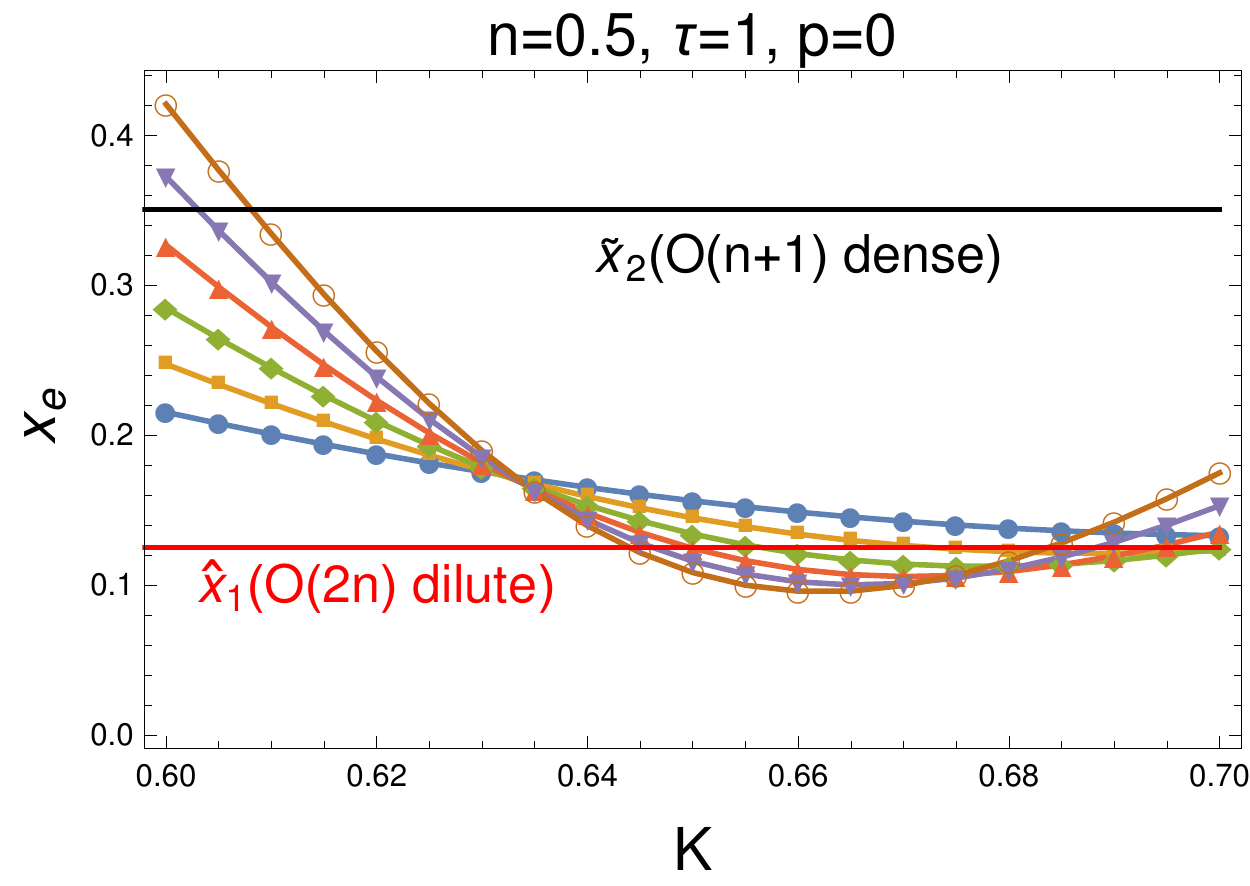}
\includegraphics[scale=0.5]{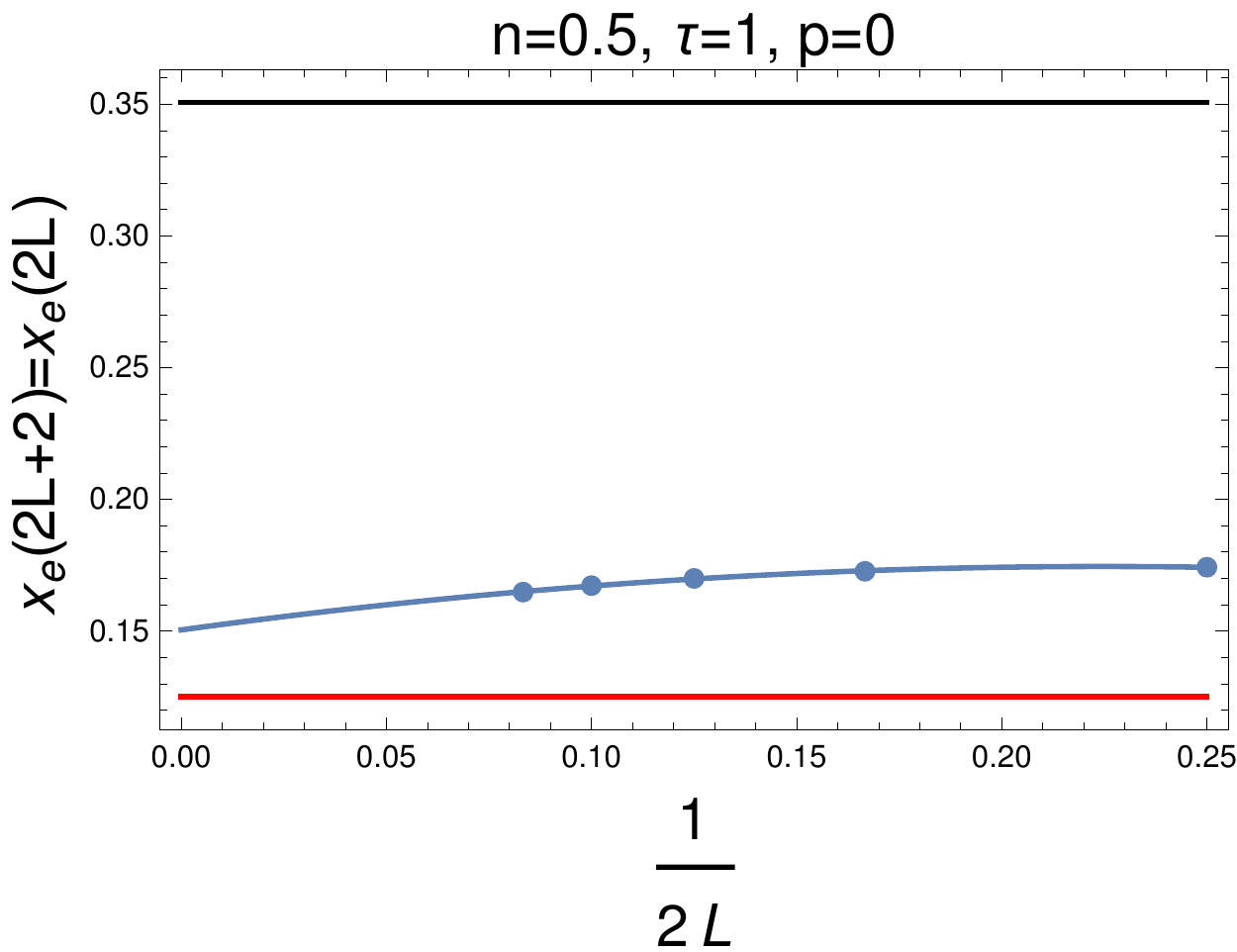}
\end{center}
\caption{Watermelon exponent $x_{\rm e} = x_{\rm o}$ across the $n=\frac12$ dilute to dense transition, at $\tau=1$ (periodic boundary conditions). We plot for comparison the exponent $\widetilde{x}_2$ for the dense {\rm O}($n+1$) model (black line), and the exponent $\hat{x}_1 = \frac{1}{8}$ for the dilute {\rm O}($2n$) model (red line). The finite-size analysis in the right panel proceeds in analogy with the preceeding figures. It shows that as $L \to \infty$ the exponent decreases, very slowly, presumably towards the dilute {\rm O}($2n$) value.}
\label{fig:tau1_xe_n05}
  \end{figure}
As shown in figure~\ref{fig:tau1_xe_n05}, where the exponent $x_{\rm e} = x_{\rm o}$ is studied, the convergence towards the expected $\hat{x}_1$ is, at best, very slow.
The analysis is based on the crossings near $K \approx 0.633$; the figure exhibits another set of less neat crossings around $K \approx 0.69$, but their physical relevance
can easily be discarded by examining the central charge. 
The low quality  of the convergence originates, we believe, in the fact that  for $n=1$, the {\rm O}($2n=2$) model has, in fact,  a {\sl line of critical points}, along which the exponents are expected to vary continuously. Let us now examine this scenario. 

\subsection{The special case $n=1$}
\label{sec:n1}

\begin{figure}
\begin{center}
\includegraphics[scale=0.65]{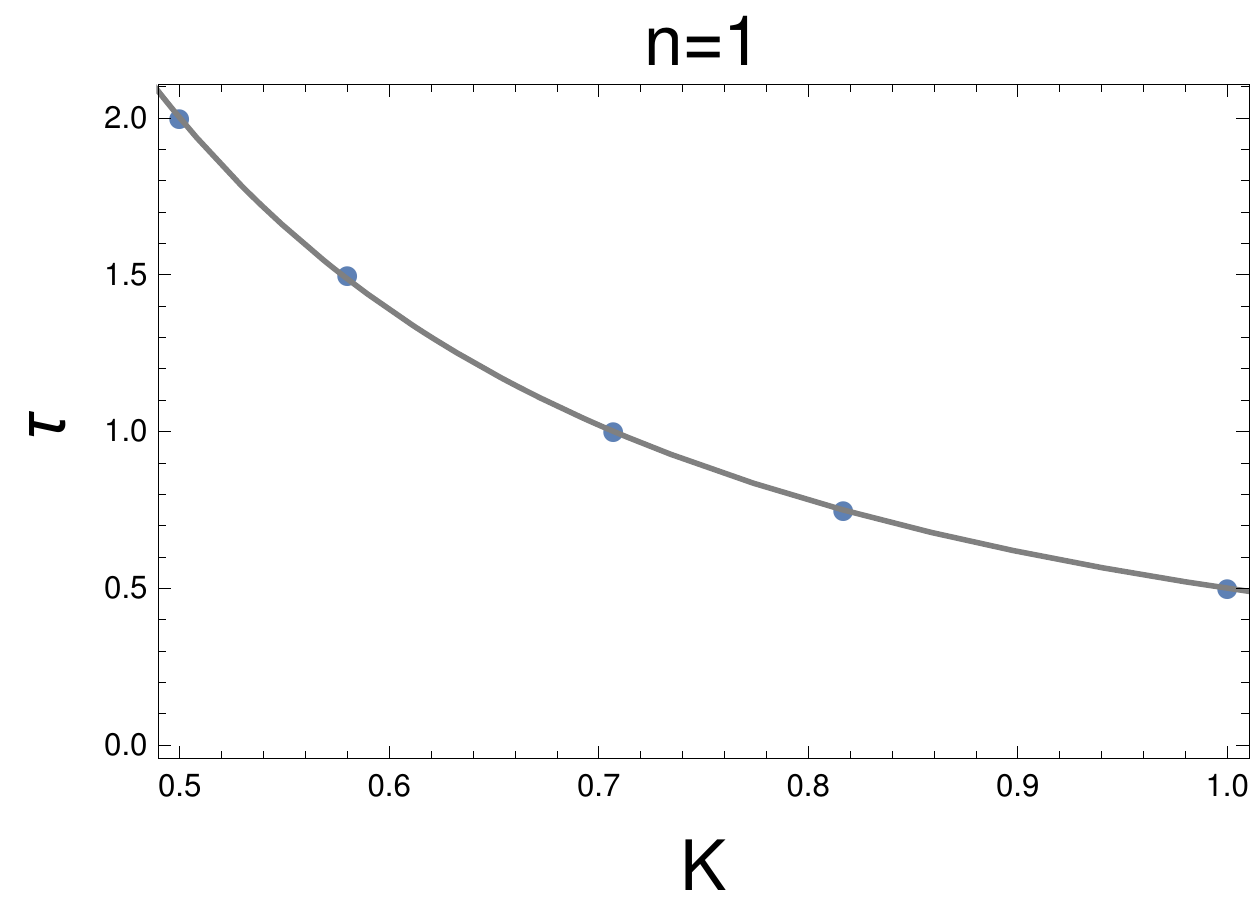}
\end{center}
\caption{
Location of the critical point $(K_c,\tau_c)$ in the $n=1$ plane, estimated numerically for several values of $\tau$ (blue dots). We plotted in comparison the line $\tau K^2 = \frac{1}{2}$, which indeed seems to parameterise the line of critical points.}                            
\label{fig:n1Kctau}
  \end{figure}

Our numerical determination of the critical line $K_{\rm c}(\tau)$ in the plane $n=1$ is shown in figure \ref{fig:n1Kctau}.
We already know analytically that it contains the ${\rm SU}(1+1)$ point at $(K,\tau) = (\frac12,2)$. The figure gives
convincing evidence that $(K_{\rm c})^2 = \frac{1}{2\tau}$ for any $\tau \in (0,2]$,
and we shall show analytically below that this is indeed the true answer.
Moreover, we shall show that the critical exponents vary continuously with $\tau$,
and give the exact values for a subset of the exponents along the critical line.

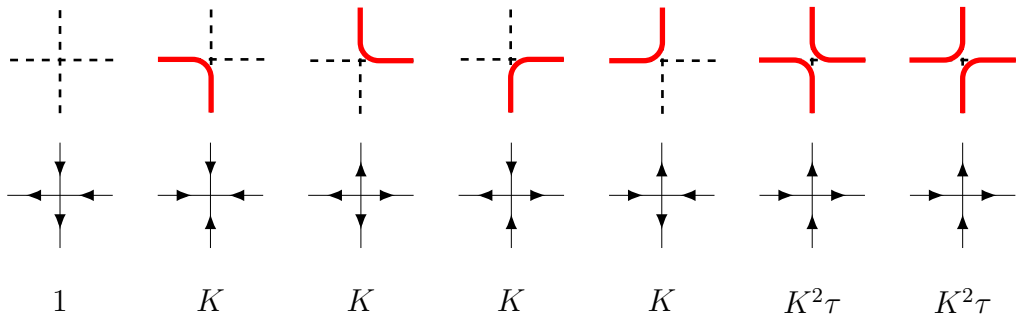
\begin{figure}
\begin{center}
\begin{tikzpicture}
\node at (0,0) {\RI};
\node at (0,-1.8) {\RIeightv};
\node at (0,-3.2) {$1$};
\begin{scope}[shift={(2,0)}]
\node at (0,0) {\RII};
\node at (0,-1.8) {\RIIeightv};
\node at (0,-3.2) {$K$};
\end{scope}
\begin{scope}[shift={(4,0)}]
\node at (0,0) {\RIII};
\node at (0,-1.8) {\RIIIeightv};
\node at (0,-3.2) {$K$};
\end{scope}
\begin{scope}[shift={(6,0)}]
\node at (0,0) {\RIV};
\node at (0,-1.8) {\RIVeightv};
\node at (0,-3.2) {$K$};
\end{scope}
\begin{scope}[shift={(8,0)}]
\node at (0,0) {\RV};
\node at (0,-1.8) {\RVeightv};
\node at (0,-3.2) {$K$};
\end{scope}
\begin{scope}[shift={(10,0)}]
\node at (0,0) {\RVIII};
\node at (0,-1.8) {\RVIIIeightv};
\node at (0,-3.2) {$K^2 \tau$};
\end{scope}
\begin{scope}[shift={(12,0)}]
\node at (0,0) {\RIX};
\node at (0,-1.8) {\RIXeightv};
\node at (0,-3.2) {$K^2 \tau$};
\end{scope}
\end{tikzpicture}
\end{center}
\caption{
Mapping from the dilute model in the $n=1$ plane onto the eight-vertex model. Occupied (resp.\ empty) horizontal edges correspond
to an arrow towards the right (resp. left), while occupied (resp.\ empty) vertical edges correspond to an up-pointing (resp.\ down-pointing)
arrow. The corresponding Boltzmann weights are also indicated. The two omitted eight-vertex configurations have zero weight.}                            
\label{fig:mapping8v}
\end{figure}

First, we notice that the model for $n=1$ can be mapped onto an eight-vertex model%
\footnote{This mapping and the subsequent mapping to a six-vertex model were already discussed in \cite{FuGuoBlote13},
but only for the particular case $(K,\tau)=(\frac{1}{\sqrt{2}},1)$.}
by assigning arrows to the edges as explained in figure \ref{fig:mapping8v}; for convenience we have rotated the
vertices by $45^\circ$ in the figure.

The last two pictures correspond to the same eight-vertex configuration, with total weight $2 K^2 \tau$. 
There is thus a total of six vertices out of the eight possible that have a non-zero weight.

Any vertex model on a bipartite lattice is invariant under a gauge transformation consisting in reversing the arrows on
the South and East edges (resp.\ the North and West edges) on the even (resp.\ odd) sublattice. We now apply this
transformation to the eight-vertex model in figure~\ref{fig:mapping8v}. The result is that we recover a staggered six-vertex
model, where the weights on the even sublattice read (in the standard notation \cite{Baxter_book})
\begin{equation}
 (\omega_1,\omega_2,\omega_3,\omega_4,\omega_5,\omega_6) = (K,K,K,K,1,2 K^2 \tau) \,,
\end{equation}
while those on the odd sublattice are
\begin{equation}
 (\omega_1',\omega_2',\omega_3',\omega_4',\omega_5',\omega_6') = (K,K,K,K,2 K^2 \tau,1) \,,
\end{equation}
We recover a homogeneous, and hence solvable, six-vertex model if and only if $\omega_i = \omega_i'$ for
all $i$, that is, $2 K^2 \tau = 1$. The corresponding anisotropy parameter is then (still in the usual notation \cite{Baxter_book})
\begin{equation}
 \Delta = \frac{a^2+b^2-c^2}{2 ab} = 1-\tau \,.
\end{equation}
When $\tau$ varies between $0$ and $2$, $\Delta$ varies between $1$ and $-1$, that is, we cover the whole critical line
of the six-vertex model. For $\tau > 2$ the model remains solvable, but since $\Delta < -1$ it belongs to a
non-critical phase whose correlation length decreases monotonically with $\tau$ \cite{Baxter_book}.

We note that the gauge transformation preserves the periodic boundary condition, so the six-vertex model enjoys
purely periodic (not twisted) boundary conditions. The central charge in the critical regime $\tau \in (0,2]$ is therefore
that of a free bosonic field, namely $c=1$ all along the line. This is indeed what we find numerically. 

The conformal dimensions can be written in the Coulomb gas setup as 
\begin{equation}
 x_{E,M} = \frac{E^2}{2\bar{g}} + \bar{g}\frac{M^2}{2} \,,
\end{equation}
where $E$ and $M$ are the so-called electric and magnetic charges (the latter is related to the six-vertex magnetisation, $M=S_z$).
We have here parameterised
\begin{equation}
 \Delta = \cos \pi \bar{g} \,,
\end{equation}
using now a bar ($\bar{\ }$) to distinguish the present theory from the two other Coulomb gases used so far.

\begin{figure}
\begin{center}
\includegraphics[scale=0.85]{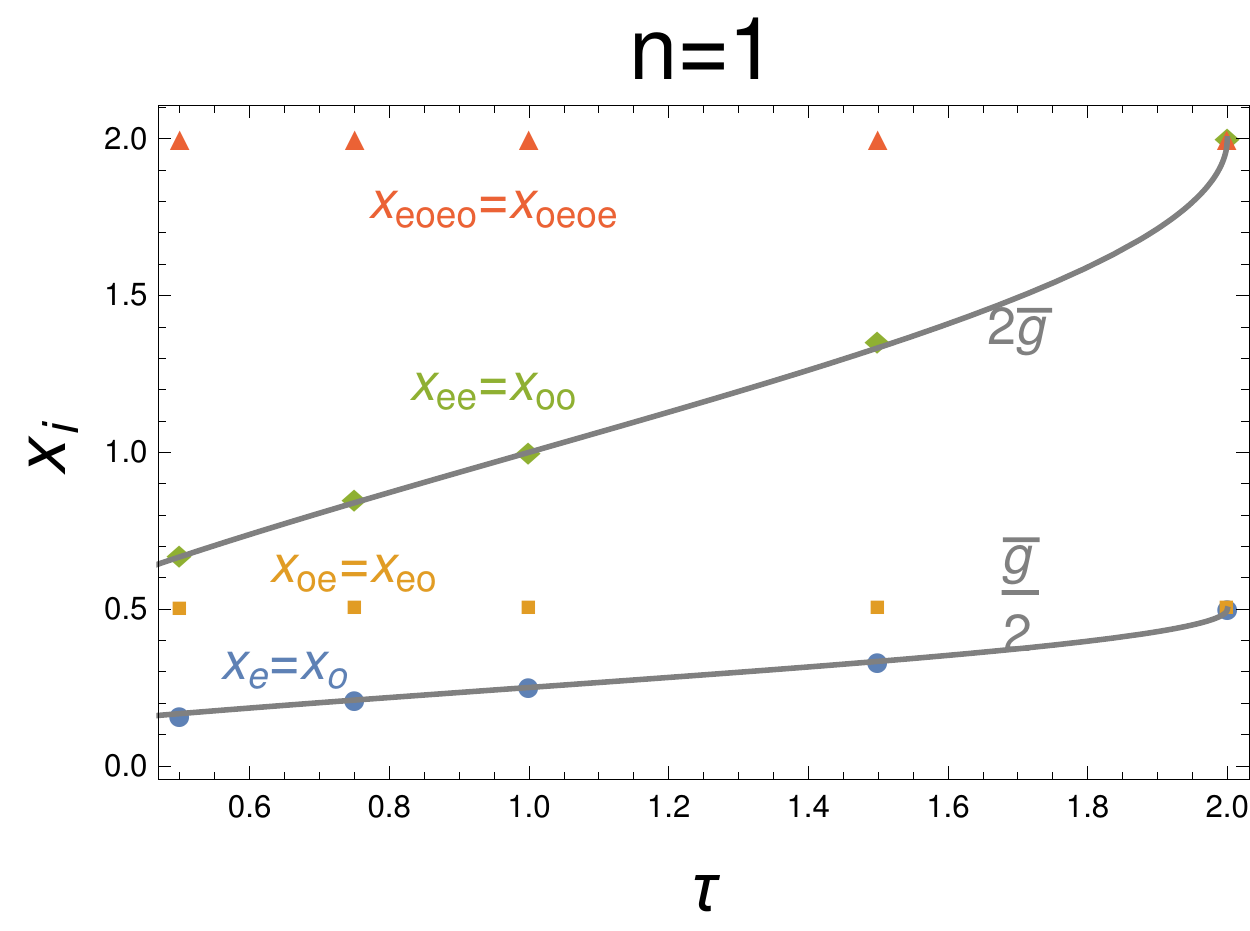}
\end{center}
\caption{
Watermelon exponents of the dilute model along the critical line for $n=1$ (periodic boundary conditions). We have plotted in comparison the expressions (\ref{xl:6v}) for $\ell=1$ and $\ell=2$, with which the exponents with pure parity labels show perfect agreement. Meanwhile, the exponents with purely alternating parity labels are constant, namely equal to their value at the ${\rm SU}(2)$ point, all along the critical line. 
}                                 
\label{fig:n1xi}
\end{figure}

Let us examine the leading eigenvalues in each sector of the dilute oriented model in the different sectors ($S_z = 0,\pm 1,\ldots,\pm L$) of the six-vertex model. For small sizes, we check the following:
\begin{eqnarray}
0 & \longrightarrow& M = 0, E=0 \,; \nonumber \\
{\rm e}={\rm o} &\longrightarrow& M= \pm 1, E=0 \,; \nonumber \\
{\rm ee}={\rm oo} &\longrightarrow& M = \pm 2, E=0 \,; \nonumber \\
{\rm eee}={\rm ooo} &\longrightarrow& M = \pm 3, E=0 \,; \nonumber \\
{\rm eo}={\rm oe} &\longrightarrow& \text{absent from the six-vertex spectrum} \,; \nonumber \\
{\rm eoe}={\rm eeo}={\rm ooe}=\ldots &\longrightarrow& \text{absent from the six-vertex spectrum} \,.
\end{eqnarray}
which we interpret as follows.
First, it is easy to see from the explicit mapping how each non-contractible parity label `e' (resp `o')
results in a $-1$ (resp $+1$) contribution to  the total value of $S_z$. 
The sectors involving parity labels of only one type are in this sense 'highest weights',
and are correspondingly found to make up the leading spectrum of the six-vertex model.
So the states $x_{\rm o},x_{\rm oo},x_{\rm ooo},\ldots$ correspond to $M=1,2,3,\ldots$ and $E=0$---and
similarly for $x_{\rm e},x_{\rm ee},x_{\rm eee},\ldots$, upon changing the sign of $M$. Therefore
\begin{equation}
 x_{\rm oo\ldots (\ell \text{ times})} = g\frac{\ell^2}{2} \,,
 \label{xl:6v}
\end{equation}
a result for which we find very good numerical evidence (see figure \ref{fig:n1xi}), and which recovers in
particular the known exponents at the ${\rm SU}(1+1)$ point (which corresponds to $(K,\tau)=(\frac{1}{2},2)$,
so that $\Delta=-1$ and namely $\bar{g}=1$).

\begin{figure}
\begin{center}
\includegraphics[scale=0.6]{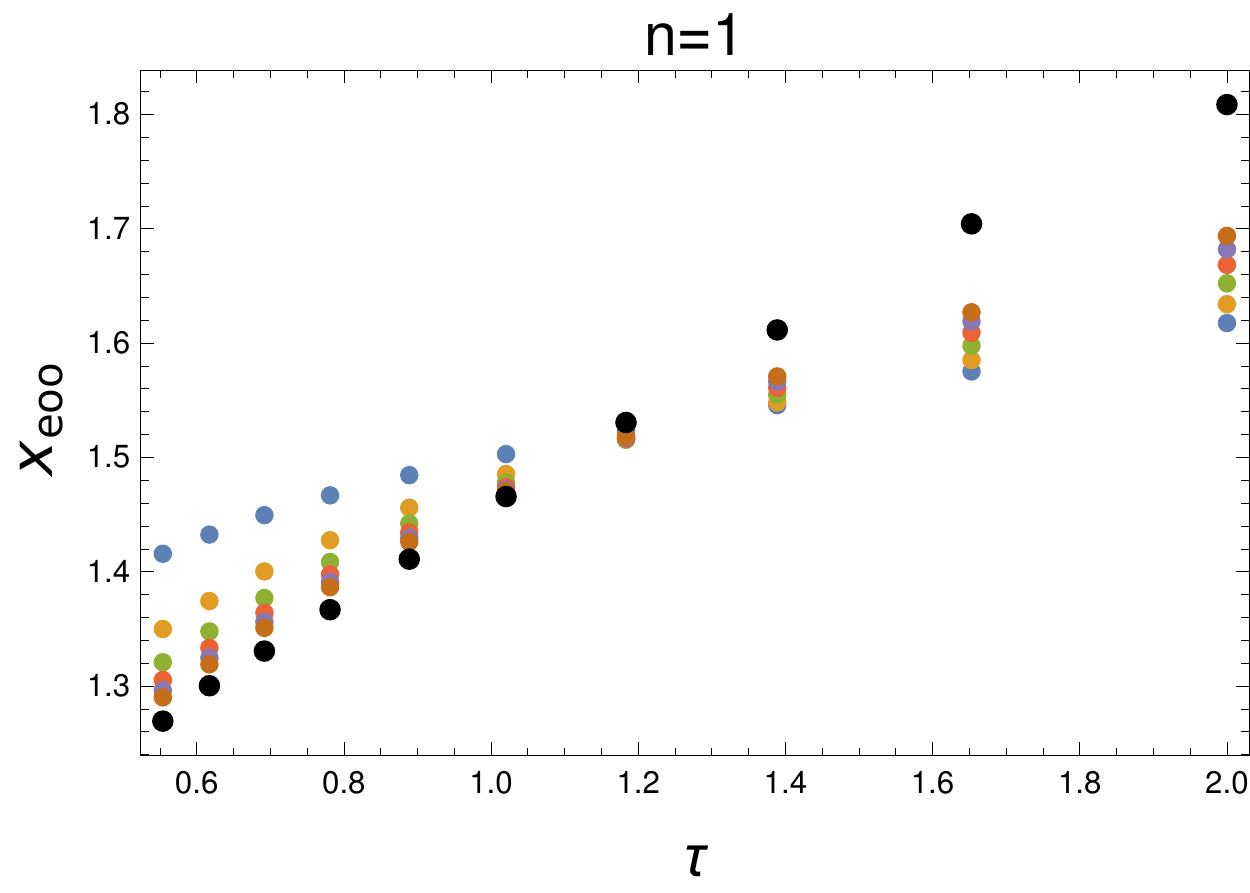}
\\
\includegraphics[scale=0.6]{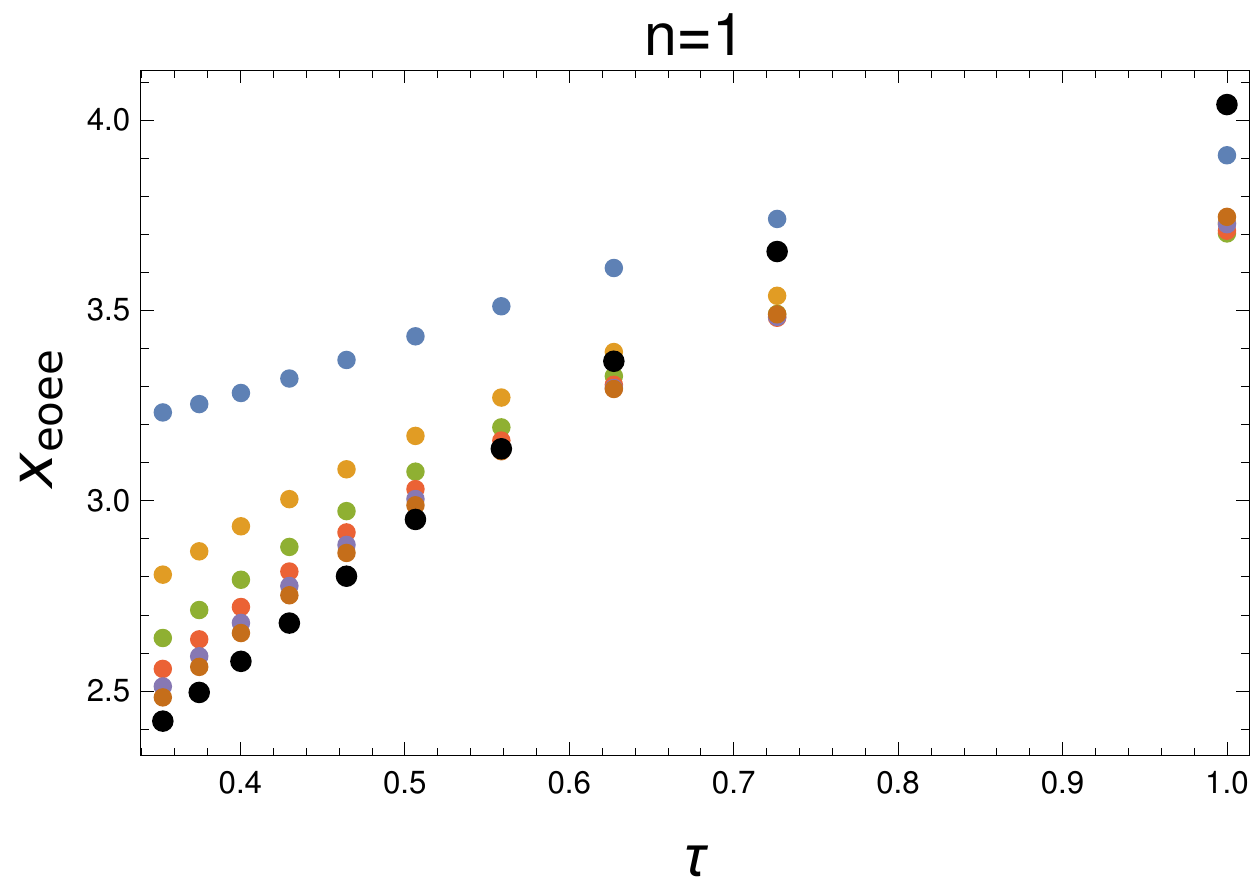}
\\
\includegraphics[scale=0.6]{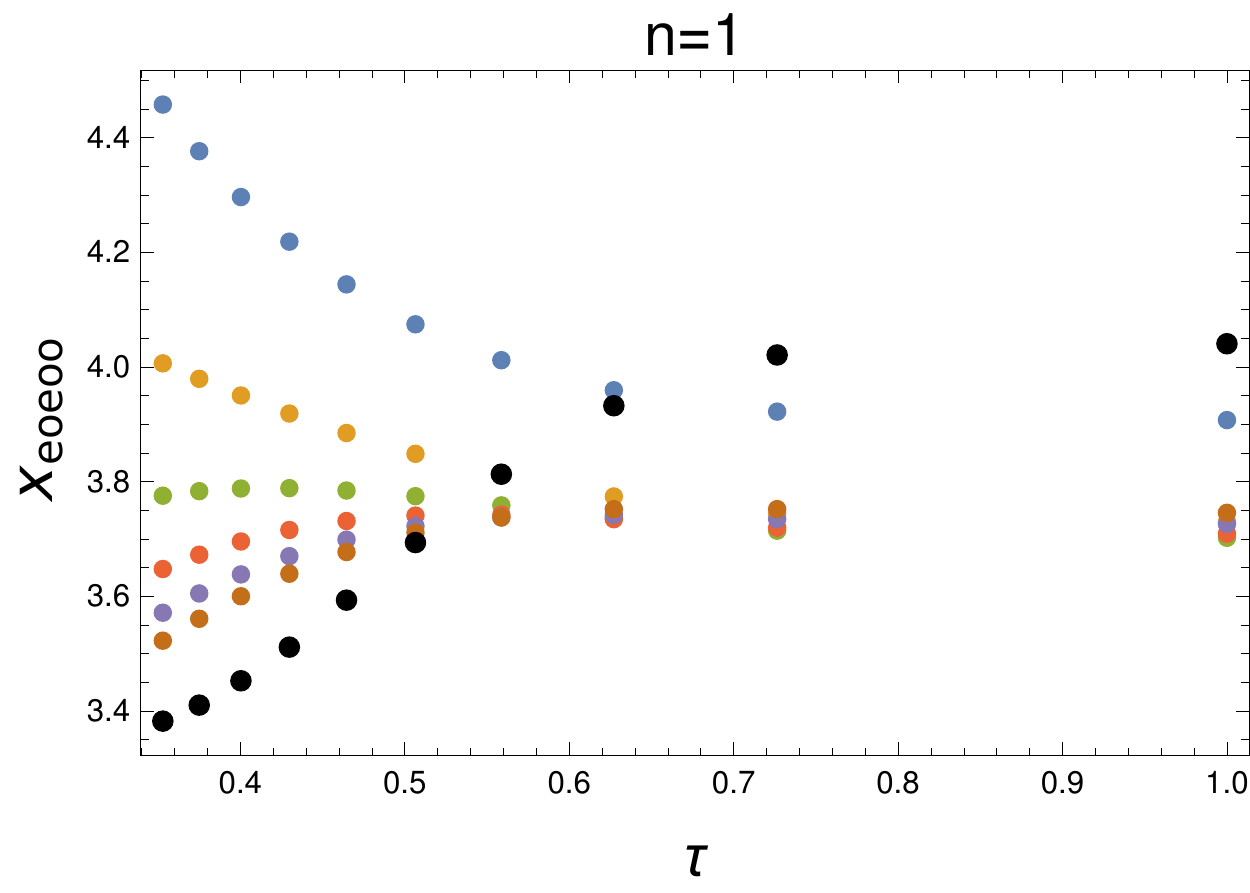}
\end{center}
\caption{
Watermelon exponents $x_{\rm eoo}$, $x_{\rm eoee}$, $x_{\rm eoeoo}$ of the dilute oriented model along the critical line for $n=1$ (periodic boundary conditions), estimated from transfer matrix diagonalisation for sizes $L=6,8,10,12,14,16$ (coloured dots). The black dots represent an extrapolation to $L\to \infty$ using a quadratic fit in $L^{-1}$. 
}                                 
\label{fig:n1xeoo}
\end{figure}

Conversely, the sectors mixing the two parity labels are found to be absent from the six-vertex spectrum, and our interpretation is that the associated operators are non-local in the six-vertex formulation. 
In practice, we observe that the exponents associated with purely alternating parity labels are constant all along the critical line, and given by their value at the ${\rm SU}(2)$ point. For instance, we find $x_{\rm eo} = x_{\rm oe} = \frac{1}{2}$, $x_{\rm eoeo} = x_{\rm oeoe} = 2$, as displayed in figure \ref{fig:n1xi}, and we conjecture that in general
\begin{equation}
 x_{\rm eoeo\ldots (\ell \text{ pairs})} = x_{\rm oeoe\ldots (\ell \text{ pairs})} = \frac{\ell^2}{2} \,.
\end{equation}
As for the remaining exponents, namely those with parity labels mixing even and odd labels in a non purely alternating way, we could not produce any analytical formula. From the results in figure \ref{fig:n1xeoo} the exponents clearly vary along the critical line, however we could not formulate a convincing conjecture.

\subsection{The special case $n=0$}

The case $n=0$ is very special, since formally ${\rm O}(2n)\equiv {\rm O}(n)$. This means that, starting with  dilute oriented loop model where loops get vanishing fugacity, we are able to reach the same universality class as the ordinary dilute model---that is, the model which does not respect the lattice orientation. In other words, the universality class on the red line of figure (\ref{fig:phasediag}) is the usual dilute ${\rm O}(0)$ universality class. Meanwhile, the critical point (violet dot) is in the universality class of the theta point as identified in \cite{DuplantierSaleur89}. 

\section{Conclusion}
\label{sec:disc}

The main result of this paper is that, while the ordinary dilute loop model with fugacity $n$ per loop has a critical point in the {\rm O}($n$) universality class, the dilute oriented loop model has a critical point in the {\rm O}($2n$) universality class. 
There are, associated with this observation, many interesting algebraic  as well as phenomenological aspects which deserve further study. In particular, note that the dilute oriented model is an special case of a more general loop model---namely the usual, non-oriented loop model
on the square lattice \cite{BloteNienhuis89}---that possesses and intriguing phase diagram whose features are not all understood at the moment, despite some recent progress \cite{VJSa22,VJSpoly}. 

Of particular importance is the behaviour for $n=0$: only in this case are the {\rm O}($n$) and {\rm O}($2n$) universality classes identical, and the underlying lattice orientation mostly irrelevant---at least in the continuum limit. In \cite{IFC11}, a loop model similar (albeit with more complicated rules, involving in particular two loop colours) to the one discussed here was extended from a definition compatible with the lattice orientation to one that is not, and it was argued that the universality class is not modified upon breaking the lattice orientation. This is not at all an obvious result, and most likely holds only for special values of the loop fugacity, just like in the model we studied in this paper.  This is an aspect we will discuss more elsewhere. 

Meanwhile, the reformulation of  the ${\rm SU}(n+1)$ model as a dilute oriented model with $S{\rm U}(n)$ symmetry turns out to be an important step in the study of loop reformulations of critical points between universality classes of topological insulators, an aspect which we will also explore in more detail  elsewhere.

\ack

We thank Sergio Caracciolo and Bernard Nienhuis for discussions, and Adam Nahum for comments on the manuscript.
Support from the Agence Nationale de la Recherche (grant ANR-10-BLAN-0414: DIME)
and the Institut Universitaire de France is gratefully acknowledged. 

\section*{References}
\bibliographystyle{iopart-num}
\bibliography{VJS}

\end{document}